%% file: cascades.tex
\documentclass[letterpaper,twocolumn,10pt]{article}
\usepackage{usenix2019_v3}
\usepackage{stfloats}

\usepackage{array}
\usepackage{booktabs}
\usepackage{subcaption}
\usepackage{xspace}
\usepackage{amsmath}
\usepackage{listings}
\usepackage{fancyvrb}
\usepackage{xcolor}
\usepackage{graphicx}
\usepackage{enumitem}
\usepackage{comment}
\usepackage{tikz}
\usetikzlibrary{positioning, arrows.meta, shapes.geometric, calc}
\definecolor{indigo}{HTML}{4B0082}
\definecolor{emerald}{HTML}{50C878}
\usepackage[most]{tcolorbox}
\usepackage{algorithm}
\usepackage{algpseudocode}
\newcommand{\To}{\textbf{to} }

\newcommand{\AlgNot}{\textbf{not} }

\makeatletter
\renewcommand\section{\@startsection {section}{1}{\z@}%
  {-2.2ex plus -0.6ex minus -.2ex}%
  {1.1ex plus .1ex}%
  {\reset@font\large\bf}}
\renewcommand\subsection{\@startsection{subsection}{2}{\z@}%
  {-1.8ex plus -0.5ex minus -.2ex}%
  {0.7ex plus .1ex}%
  {\normalfont\large\bfseries}}
\makeatother

\newtcbox{\challengebadge}{
  on line,
  arc=2pt,
  colback=gray!20!white,
  colframe=black!70,
  boxrule=0.6pt,
  boxsep=0pt,
  left=4pt, right=4pt, top=2.5pt, bottom=2.5pt,
  fontupper=\bfseries\sffamily\small
}

\newcommand{\systemname}{LOGOS\xspace}

\makeatletter
\def\@maketitle{\newpage
 \begin{center}%
  {\Large\bf \@title \par}%
  \vskip 0.08in
  {\large\it
   \lineskip .3em
   \begin{tabular}[t]{c}\@author
   \end{tabular}\par}%
 \end{center}%
}
\makeatother

\begin{document}

\title{\Large \bf From Alert Floods to Precedence Forests: Zero-Prior-Knowledge Incident Triage with LOGOS}

\author{
{\rm Radhika Niranjan Mysore}
}

\maketitle

\begin{abstract}
\input{sections/abstract.tex}
\end{abstract}

\input{sections/intro}
\input{sections/background}
\input{sections/upgrade}
\input{sections/design}
\input{sections/impl}
\input{sections/eval}
\input{sections/related}
\input{sections/limits}
\input{sections/conc}

\bibliographystyle{plain}
\bibliography{refs}

\end{document}

%% file: sections/abstract.tex
Commercial observability platforms rely on domain artifacts like distributed traces, topology maps, and baseline metrics. However, when troubleshooting proprietary software, enterprise operators are left with only raw, unannotated text logs. We explore the extreme boundary of log-only diagnosis: \textit{To what extent can we isolate failure propagation using strictly raw text logs?} We present \systemname{}, an unsupervised system that exploits entity-event co-occurrence and temporal precedence to collapse millions of raw log lines into a compact precedence forest. Evaluated across 25 production enterprise outages and 12 open-source issues, \systemname{} operates with zero prior knowledge—requiring no seed queries, observed symptoms, or pre-defined incident boundaries. In a median wall-time of 4.5~minutes, \systemname{} eliminates a median $99.8\%$ of background noise, achieves $0.76$ mean recall, and detects failure cascades with a 16-hour median diagnosis-verified lead time—consolidating alert floods $124\times$ to enable $80\%$ enterprise ($100\%$ open-source) zero-shot LLM root-cause accuracy.

%% file: sections/intro.tex
\section{Introduction}
\label{sec:intro}
Modern observability platforms~\cite{splunk, datadog, dynatrace} rely on rich domain artifacts---dependency maps, distributed traces, and execution baselines---that are absent in black-box enterprise deployments. Enterprise IT administrators lack the source code, vendor-specific domain expertise, and incident history required to inspect proprietary vendor binaries, while vendor engineering teams lack visibility into the customer's private deployment state until an outage surfaces~\cite{xu2026aida,he2022empirical,skynet}. This creates a \textit{double bind}: troubleshooting guides and runbooks quickly become stale~\cite{he2022empirical, ghosh2022prod, pc24livedebug, shetty2022autotsg}, and emergent cross-vendor failures fall outside any single vendor's source code or documentation. Consequently, enterprise debugging remains overwhelmingly manual~\cite{pc24livedebug}. While recent AIOps pipelines~\cite{jiang2020mitigate, shetty2022autotsg, chen2024rcacopilot, wang2024rcaagent, zhang2025tamo, xu2026aida, ren2026xihe} attempt to automate triage, they assume curated playbooks or historical incident archives that enterprise environments do not possess.

Because raw text logs remain the only universally accessible artifact spanning all vendor boundaries~\cite{he2022empirical} (\S\ref{sec:background}), this work explores the extreme end of the spectrum: \textit{To what extent can we bound the diagnostic search space using strictly raw text logs?} By establishing this lower bound---without requiring source code, distributed traces, or healthy execution baselines---log-only analysis empowers enterprise administrators to troubleshoot complex multi-vendor incidents as far as possible on their own. When severe issues exceed in-house resolution and require escalation to vendor support, this bounded representation provides a distilled, cross-vendor failure context—replacing unguided log grepping with an empirical baseline upon which enterprise operators, vendor engineers, and future AIOps pipelines can build.

Existing log-only diagnostic studies~\cite{xu2009detecting, lin2016logcluster, zhang2021onion, du2017deeplog, meng2019loganomaly, zhang2019logrobust, li2020swisslog, guo2021logbert, han2023loggpt, jia2023rtlog} rely on clean, stationary healthy baselines. However, enterprise infrastructure rarely yields such baselines: healthy telemetry is absent during new deployments and upgrades, and otherwise non-stationary, sparse, or unrecorded during normal operation~\cite{kandula2009netmedic, he2022empirical, jia2023rtlog}. Consequently, how far raw text logs alone can bound the \textit{enterprise} diagnostic search space without healthy execution baselines remains an open question.

\noindent\textbf{Our contribution.} In this paper, we present \systemname{} to answer this question. \systemname{} is a log-only analysis system that bounds and orders the diagnostic search space by inferring approximate fault propagation across vendor infrastructure subsystems. For example, when applied to virtual infrastructure logs, \systemname{} isolates latent faults like unpersisted configuration changes and infers how they propagate to disrupt virtual network bring-up.

The goal of \systemname{} is to serve as a fully automated \textit{first step} before an enterprise administrator begins manual triage. Therefore, \systemname{} must trace faults through raw telemetry \textit{without relying on external inputs}.

To \textbf{bound the search space}, \systemname{} eliminates background noise by leveraging system and runtime entity identifiers embedded in log text to link \textit{related events} (events sharing entities) and \textit{related entities} (entities \textit{co-occurring} across events). Together, these form a bipartite entity-event graph; computing the transitive closure over this graph isolates related events from unrelated background noise (e.g., as illustrated by \texttt{config\_store} in Listing~\ref{lst:real_incident_timeline_masked}). Within this graph, higher-severity events (e.g., \texttt{[WARN]}, \texttt{[ERROR]}) denote faults, while participating identifiers represent \textit{impacted entities}—either originating faults or downstream victims.

To \textbf{order the search space}, \systemname{} sorts impacted entities in this isolated graph by their earliest faults. However, pure chronological sorting alone does not reveal how \textit{faults propagate} between entities that do not directly share a log event. To establish fault precedence from log data, \systemname{} connects each impacted entity to its nearest earlier co-occurring entity in a parent-child relationship, reconstructing the \textit{failure propagation sequence}. 

This log-only reconstruction is challenging for two primary reasons. First, \textit{identifying faults in logs is hard.} Enterprise logs do not encode the correct severity of events. In virtual infrastructure logs, third-party vendor subsystem events are logged verbatim without assigned severities, because they do not share common logging conventions or error codes, leaving them at risk of being misclassified as benign \texttt{[INFO]} events. While determining exact event severity without domain knowledge is impossible, \systemname{} captures these faults by elevating central subsystem logs that wrap such third-party events directly to fault candidates. Additionally, it extracts embedded severity keywords from free-form log text; for instance, the line \texttt{[t0 - 37m] [INFO] management\_node: Ignoring config\_parse\_exception[1046]: Not a non-singleton object: key alert} (Listing~\ref{lst:real_incident_timeline_masked}) is promoted based on its embedded \texttt{alert} keyword—a heuristic we call severity elevation.

Second, \textit{identifying the boundary of failure propagation, especially its origin, is hard.} Entity correlation alone cannot distinguish originating faulty entities from downstream victims. Blindly correlating all impacted entities across an entire multi-day log bundle inflates the search space as victims co-occur with unrelated background events. To isolate the failure origin without domain knowledge, \systemname{} partitions the log timeline into uniform windows and flags fault propagation onset at the first window where correlated faults span two or more distinct subsystems—marking the start of a multi-subsystem \textit{cascade}. Starting from this onset window, \systemname{} correlates faults across consecutive windows, constructing a bipartite entity-event graph and computing its transitive closure to isolate independent subgraphs. \systemname{} tracks the maximal connected component (the primary propagation cascade detected) and expands it until a subsequent window adds no new faults, at which point graph construction terminates. We call this process progressive onset tracking.

\systemname{} organizes the impacted entities in this primary cascade into parent-child relationships to reconstruct the failure propagation sequence. However, this construction is confounded by entities that occur ubiquitously in logs—such as global session identifiers or cluster-wide task handles used by management subsystems—because they co-occur with nearly all other entities. Lacking domain context, \systemname{} would treat these handles as universal parent hubs, causing them to falsely appear as the root of all other faults. To prevent this structural distortion, \systemname{} identifies and filters out ubiquitous entities prior to tree construction—a technique we call hub-node pruning. Removing these hub nodes partitions the cascade into a precedence forest of distinct propagation trees suited for direct human triage or AIOps synthesis.

\noindent To summarize, this work makes three core contributions:

\noindent\textbf{System Design.} We introduce \systemname{}, an unsupervised, domain-agnostic log system that bounds the enterprise diagnostic search space in linear $O(M)$ memory, where $M$ is the number of entity--event incidences (edges of the bipartite graph) in the log bundle, without requiring source code, distributed traces, or execution baselines.

\noindent\textbf{Early Detection.} Progressive onset tracking isolates developing failures hours before catastrophic outages—achieving a 16-hour median diagnosis-verified lead time—while collapsing raw alerts into precedence forests ($124\times$ median reduction).

\noindent\textbf{Empirical Validation.} Across 25 enterprise outages and 12 open-source incidents, \systemname{} achieves $0.76$ mean recall and a median $99.8\%$ noise reduction in a median wall-time of 4.5~minutes, outperforming $O(M^2)$ graph centrality baselines. Zero-shot LLM (Gemini) evaluation yields 80\% enterprise and 100\% open-source root-cause accuracy—establishing an empirical log-only lower bound upon which LLM reasoning and AIOps pipelines can build.

%% file: sections/background.tex
\section{Background and Motivation}
\label{sec:background}

Log analysis is the primary diagnostic method for complex enterprise outages. Across 100 severe virtual infrastructure incidents we studied, spanning distributed enterprise datacenters across compute, storage, networking, and control planes, 56 were root-caused using logs alone, 7 required fine-grained metrics, 7 required memory core dumps, and 8 required source code analysis (primarily locking and resource contention). The remaining 22 were resolved out-of-band via upgrades or reconfigurations—a common outcome for complex enterprise issues~\cite{gunawisocc2016, ghosh2022prod}. Diagnosing the 56 log-centric issues was an inherently manual and iterative process spanning vendor support and engineering teams, averaging $\sim$33 days in calendar resolution time. \systemname{} directly targets the empirical cost of unguided log search by eliminating a median $99.8\%$ of background noise in a median compute wall-time of 4.5~minutes (\S\ref{sec:onset}, \S\ref{sec:overheads})—narrowing, rather than replacing, this 33-day manual investigation. From the 56 log-centric issues, we selected 25 log-diagnosable issues for evaluating \systemname{} based on complete log bundle availability and verified ground truth.

The 25 issues highlight how faults cross subsystem and vendor boundaries in complex enterprise outages. Table~\ref{tab:incident_categories_collapsed} summarizes the symptom and root-cause categories across these incidents, revealing two key characteristics. First, third-party vendor issues account for nearly two-thirds of all incidents (16 of 25). This concentration reflects the operational pain point resulting from emergent cross-vendor interactions where administrators lack source code or runbooks. Second, root causes rarely reside in the same subsystem as the observed symptom~\cite{zhang_csnake26} and are often logged days before the final outage~\cite{huang2017gray}. Consequently, diagnosing these incidents requires analyzing multi-day logs across several decoupled subsystems to trace terminal symptoms back to their originating root cause.

\begin{table}[htb]
\centering
\small
\caption{\textbf{Symptom and Root Cause Category breakdown across the 25 enterprise evaluation incidents.}}
\label{tab:incident_categories_collapsed}
\begin{tabular}{@{}llr@{}}
\toprule
\textbf{Symptom Category} & \textbf{Root Cause Category} & \textbf{Count} \\ \midrule
Compute & Third-party storage & 2 \\ \addlinespace
Control plane & Compute / Config & 1 \\
Control plane & Network / Config & 1 \\
Control plane & Storage / Config & 1 \\ \addlinespace
Management plane & Config & 3 \\
Management plane & Memory & 2 \\
Management plane & Third-party management plane & 1 \\ \addlinespace
Memory & Third-party config & 1 \\ \addlinespace
Network & Third-party config & 1 \\
Network & Memory / Config & 1 \\ \addlinespace
Storage & Third-party storage & 9 \\
Storage & Third-party management plane & 1 \\
Storage & Third-party network & 1 \\ \midrule
\textbf{Total Incidents} & & \textbf{25} \\ \bottomrule
\end{tabular}
\end{table}

While simple frequency and severity ranking heuristics capture root-cause entities in nearly half of these incidents (12 of 25), flat entity lists fail to reconstruct the cross-subsystem failure propagation required for diagnosis, detailed next.

\noindent\textbf{Naive Entity-Ranking Heuristics.}
To evaluate whether naive entity-ranking heuristics can resolve these incidents, we tested simple frequency and severity baselines on candidate faults isolated via severity elevation. While these rankings capture the root-cause entity in nearly half the incidents (12 of 25), a flat ranked list fails to reveal how the failure propagated across subsystem boundaries~\cite{xu2026aida, ren2026xihe}. Consequently, operators are still forced to manually investigate each entity as an isolated starting point, leaving the diagnostic search space unreduced. Table~\ref{tab:heuristic_evaluation} breaks down the number of incidents each heuristic surfaced.

\begin{table}[htb]
\centering
\small
\caption{\textbf{Evaluation of naive entity-ranking heuristics across 25 enterprise incidents.} Shows the number of incidents where each heuristic successfully surfaced ground-truth root-cause entities within its Top-10 list. Combined, at least one heuristic worked in 12 of 25 incidents ($48.0\%$).}
\label{tab:heuristic_evaluation}
\resizebox{\linewidth}{!}{%
\begin{tabular}{@{}lr@{}}
\toprule
\textbf{Heuristic Description} & \textbf{Caught} \\ \midrule
Top entities by raw cross-subsystem fault count & 7 \\
Top entities by distinct cross-subsystem fault spread & 2 \\
Top entities by cross-subsystem fault severity & 0 \\
Top-10 entities per subsystem by fault count & 12 \\
Top-10 entities per subsystem by fault severity & 0 \\
Low-frequency ($\le 15$) application-level entities & 0 \\ \midrule
\textbf{Combined Union (Any Heuristic Surfaced Root Cause)} & \textbf{12} \\
\textbf{Missed (No Heuristic Surfaced Root Cause)} & \textbf{13} \\ \bottomrule
\end{tabular}%
}
\end{table}

\noindent\textbf{Enterprise Telemetry.}
When enterprise incident tickets are filed, administrators upload diagnostic log bundles spanning retention windows from days to weeks. Bundles encompass telemetry across control, management, and data planes, organized in directories representing clusters, with separate log files for specialized subsystems (e.g., host management daemons, hypervisor kernels, virtual storage drivers). While enterprise bundles contain heterogeneous artifacts such as memory crash dumps and metric time-series, \systemname{} operates strictly on raw text logs.

Each log line comprises four primary fields: a timestamp, emitting subsystem, native severity level (e.g., \texttt{INFO}, \texttt{ERROR}), and free-form log text. Because native log levels are unreliable~\cite{he2022empirical}, \systemname{} applies severity elevation. Log fields are parsed using vendor-specific regular expressions, abstracting free-form text into parameterized \textbf{log templates} (or \textbf{events}) following standard log abstraction models~\cite{he2017drain}. This abstraction condenses millions of repetitive log lines into a compact vocabulary of event patterns—each representing a distinct logging statement in the underlying code—while isolating dynamic parameters (e.g., task handles, disk IDs) to serve as entity ``glue'' in our bipartite graph.

\noindent\textbf{Entity Co-Occurrence.}
To verify whether log lines naturally link events across subsystems, we analyzed telemetry across three platforms: 25 enterprise virtual infrastructure bundles, 4 OpenStack cloud datasets~\cite{openstack}, and 8 TrainTicket microservice benchmarks~\cite{trainticket}. Across all three environments, individual log lines contain multiple co-occurring identifiers ($N_{\text{line}} > 1$, averaging 2.57 to 4.37 IDs per line). This validates our core thesis: dynamic identifiers (e.g., task handles, VM UUIDs, disk IDs) naturally appear together in log text, acting as the ``glue'' that stitches cross-subsystem failure paths without manual schemas or tracing headers.

Crucially, attempting to convert these log lines into standard pairwise entity graphs used by traditional graph-based algorithms~\cite{kandula2009netmedic, wu2020microrca, bahl2007sherlock} creates dense cliques (up to $68.21\%$ overall density; Table~\ref{tab:entity_density}). In practice, this $O(M^2)$ projection bottleneck (where $M$ is the number of entity--event incidences in the bundle) causes standard centrality baselines to run out of memory on 9 of our 25 enterprise incident bundles ($36\%$). \systemname{} avoids this quadratic scaling by preserving the native bipartite structure between entities and log events, keeping computation linear ($O(M)$) and evaluating successfully across all 25 bundles.

\begin{table}[t]
\centering
\caption{\textbf{Entity Density \& Unipartite Projection Density.} $N$: bundles evaluated per platform. Individual log lines average $N_{\text{line}} > 1$ entities, forming natural bipartite edges; $N_{\text{co-occur}}$ is the mean number of entities each entity co-occurs with in a pattern. Converting these into unipartite entity-entity graphs inflates edge density ($\rho_{\text{max}}$), causing baseline graph algorithms to suffer Out-of-Memory (OOM) failures.}
\label{tab:entity_density}
\scriptsize
\begin{tabular}{@{}p{2.3cm}p{0.5cm}>{\centering\arraybackslash}p{1.0cm}>{\centering\arraybackslash}p{1.1cm}>{\centering\arraybackslash}p{1.3cm}@{}}
\toprule
\textbf{Dataset / Stack} & \textbf{$N$} & \textbf{$N_{\text{line}}$} & \textbf{$N_{\text{co-occur}}$} & \textbf{$\rho_{\text{max}}$} \\
\midrule
Virtual Infrastructure & 25 & 4.37 & 5.32 & \textbf{68.21\%} \\
OpenStack              & 4  & 2.61 & 7.58 & 56.06\% \\
TrainTicket            & 8  & 2.57 & 24.21 & 10.85\% \\
\bottomrule
\end{tabular}
\end{table}

%% file: sections/upgrade.tex
\section{Upgrade Incident Case Study}
\label{sec:case_study}
Consider a real enterprise incident where virtual machines (VMs) failed to migrate following a hypervisor upgrade. Despite state-of-the-art observability tooling and tens of gigabytes of telemetry spanning $\sim$3~years, enterprise administrators could not isolate the cause. The ticket was escalated to vendor support teams and routed to core engineering, where engineers spent over a month debugging and deploying workarounds before tracing the root cause out-of-band to a product defect that prevented host configuration updates from being persisted to disk for two years.

In this case study, the product defect itself is not explicitly recorded in the logs, but its downstream impact on the infrastructure is. \systemname{} must trace this impact through raw telemetry, isolating fault propagation from background noise \textit{without relying on external inputs}. Accordingly, it is provided strictly with the raw 20.5~million-line log bundle—requiring no starting query, observed symptom seed, or pre-defined incident boundaries.

\noindent\textbf{Incident Timeline.} Listing~\ref{lst:real_incident_timeline_masked} shows an excerpt of the log lines capturing the incident. Two years prior ($t_0 - 2\text{yr}$), a maintenance script (\texttt{backup\_cron}) marked the host as pending reboot (\texttt{pending\_reboot=1}). Expecting an imminent reboot, the configuration store (\texttt{config\_store}) \textit{silently} stopped persisting new changes to disk—a latent product defect. Because the host was never actually rebooted, all subsequent configuration updates over the next two years (including virtual switch and network group renames) existed strictly in volatile RAM. Pre-upgrade log lines at $t_0 - 116\text{h}$ and $t_0 - 37\text{m}$ recorded early warnings of this latent state drift.

When administrators rebooted the hypervisor following the upgrade at $t_0$, the host restored the stale two-year-old disk configuration ($t_0 + 12\text{m}11\text{s}$ to $t_0 + 12\text{m}11.3\text{s}$). Driven by this stale state, the switch restore daemon (\texttt{virtual\_switch\_info}) re-applied outdated port bindings, triggering network configuration errors that ultimately caused VM migration failures. Concurrently, host management daemons repeatedly attempted to re-attach the failed switch endpoint (\texttt{add\_virtual\_switch\_to\_host} from $t_0$ to $t_0 + 24\text{m}$).

\begin{figure}[htbp]
\centering
\lstset{morecomment=[l]{\%},commentstyle=\itshape\color{gray}}
\begin{lstlisting}[
  basicstyle=\ttfamily\scriptsize,
  frame=single,
  breaklines=true,
  caption={\textbf{Upgrade Incident Log Excerpt (Masked).} Chronological snippet from a multi-year log snapshot. Text following \texttt{\%} denotes added annotations (event counts and context) rather than literal log output.},
  label={lst:real_incident_timeline_masked}
]
[~2y prior]  [INFO]  backup_cron:
             Workflow run with pending_reboot=1; Live-install status
             never persisted to config_store
             % reconstructed from incident report; no raw-log timestamp
[t0 - 116h]  [WARN]  host_agent:
             Skip saving virtual_net_group to: filePath empty
             % recurs ~every 5 min from here through the reboot
[t0 - 37m]  [INFO]  management_node:
             Ignoring config_parse_exception[1046]:
             Not a non-singleton object: key alert
[t0 + 0s .. +24m]  [ERROR] host_agent:
             add_virtual_switch_to_host: object not found /
             hostcfg disabled
             % recurs 13x across this window, never resolves
[t0 + 12m11s]  [ERROR] boot_init:
             config_store: Failed to connect to database
             % repeats 4x, 3ms apart
[t0 + 12m11.2s]  [INFO]  boot_init:
             config_store: load_store Start.
[t0 + 12m11.2s]  [INFO]  boot_init:
             config_store: Restore config_store from config_store_backup
[t0 + 12m11.2s]  [INFO]  boot_init:
             config_store: config_store restored successfully from backup
[t0 + 12m11.2s]  [INFO]  boot_init:
             config_store: data_file_store backup (data_file_backup) is valid
[t0 + 12m11.3s]  [INFO]  boot_init:
             config_store: data_file_store restored successfully from backup
[t0 + 12m21s]  [INFO]  boot_init:
             virtual_switch_info: Prepare to restore: virtsw_local_binding
             % excerpt only: thousands more real lines/entities not shown
\end{lstlisting}
\end{figure}

\begin{figure*}[tb]
\centering
\scriptsize
\resizebox{\textwidth}{!}{%
\begin{tikzpicture}[node distance=0.4cm and 0.25cm,
  every node/.style={draw, rounded corners, align=center, font=\tiny, minimum height=0.55cm}]

\node (e1) [fill=gray!8] {Entity:\\ \texttt{config\_parse\_exception}};
\node (e2) [right=of e1, fill=gray!8] {Entity:\\ \texttt{add\_virtual\_switch\_to\_host}};
\node (e3) [right=of e2, fill=gray!8] {Entity:\\ \texttt{config\_store}};
\node (e6) [right=1.6cm of e3, fill=gray!8] {Entity:\\ \texttt{load\_store}};
\node (e7) [right=1.8cm of e6, fill=gray!8] {Entity:\\ \texttt{config\_store\_backup}};
\node (e8) [right=1.6cm of e7, fill=gray!8] {Entity:\\ \texttt{data\_file\_store}};
\node (e9) [right=1.6cm of e8, fill=gray!8] {Entity:\\ \texttt{data\_file\_backup}};
\node (bi) [right=1.8cm of e9, very thick, fill=white] {Subsystem:\\ \texttt{boot\_init}};
\node (e4) [right=1.8cm of bi, fill=gray!8] {Entity:\\ \texttt{virtual\_switch\_info}};
\node (e5) [right=of e4, fill=gray!8] {Entity:\\ \texttt{virtsw\_local\_binding}};

\node (ev1) [below=1.5cm of e1, fill=gray!20] {Event ($t0{-}37$m)\\ ``Ignoring config\_parse\_\\ exception[1046]...''};
\node (ev2) [below=1.5cm of e2, fill=gray!20] {Event ($t0{+}0$s)\\ ``...object not found/\\ hostcfg disabled''};
\node (ev3) [below=1.5cm of e3, fill=gray!20, text width=1.5cm] {Event ($t0{+}12$m11s)\\ ``Failed to connect\\ to database''};
\node (ev3b) [below=1.5cm of e6, fill=gray!20, text width=1.5cm] {Event ($t0{+}12$m11.2s)\\ ``load\_store Start.''};
\node (ev3c) [below=1.5cm of e7, fill=gray!20, text width=1.5cm] {Event ($t0{+}12$m11.2s)\\ ``Restore config\_store\\ from config\_store\_backup''};
\node (ev3d) [below=1.5cm of e8, fill=gray!20, text width=1.5cm] {Event ($t0{+}12$m11.2s)\\{}``data\_file\_store backup\\{}(data\_file\_backup) is valid''};
\node (ev4) [below=1.5cm of e4, fill=gray!20] {Event ($t0{+}12$m21s)\\ ``Prepare to restore:\\ virtsw\_local\_binding''};

\node (hub1) [above=0.6cm of ev1, xshift=-0.5cm, draw=none, font=\tiny, align=center] {(shared hub\\ entities)};
\node (hub2) [above=0.6cm of ev2, xshift=-0.5cm, draw=none, font=\tiny, align=center] {(shared hub\\ entities)};

\draw (e1) -- (ev1) node[midway, right, draw=none, font=\tiny] {\texttt{management\_node}};
\draw (e2) -- (ev2) node[midway, right, draw=none, font=\tiny] {\texttt{host\_agent}};
\draw[dashed] (ev1) -- (hub1);
\draw[dashed] (ev2) -- (hub2);
\draw (e3) -- (ev3);
\draw (e3) -- (ev3b);
\draw (e3) -- (ev3c);
\draw (e3) -- (ev3d);
\draw (e6) -- (ev3b);
\draw (e7) -- (ev3c);
\draw (e8) -- (ev3d);
\draw (e9) -- (ev3d);
\draw (e4) -- (ev4);
\draw (e5) -- (ev4);

\draw[very thick] (bi) -- (ev3);
\draw[very thick] (bi) -- (ev3b);
\draw[very thick] (bi) -- (ev3c);
\draw[very thick] (bi) -- (ev3d);
\draw[very thick] (bi) -- (ev4);

\node (more) [right=1.0cm of e5, draw=none, dashed, font=\tiny, align=left] {$\cdots$\\ (thousands more\\ entities/events,\\ not shown)};
\draw[dashed] (e5) -- (more);

\end{tikzpicture}%
}
\caption{\textbf{Bipartite Entity-Event Subgraph Excerpt.} Edges link entity nodes (top) to log event nodes (bottom). \texttt{config\_store} connects the database failure to backup-restore events, while \texttt{boot\_init} links them to the virtual switch restore attempt. Dashed edges connect events associated with \texttt{config\_parse\_exception} and \texttt{add\_virtual\_switch\_to\_host} to the rest of the full graph via elided hub entities prior to hub-node pruning.}
\label{fig:real_bipartite_masked}
\end{figure*}
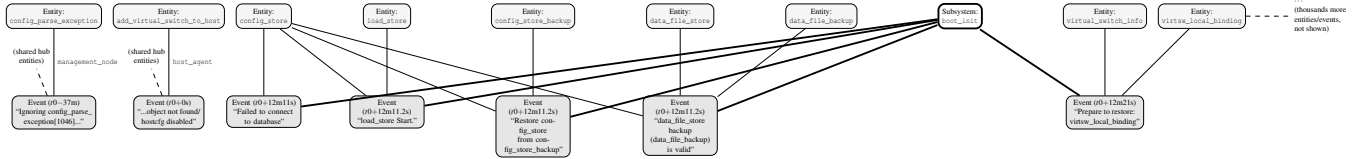

\noindent\textbf{\systemname{}'s Log Representation.} \systemname{} parses raw logs into events and entities. Severity elevation classifies all events in Listing~\ref{lst:real_incident_timeline_masked} as faults, even those carrying benign \texttt{[INFO]} tags. Specifically, \texttt{management\_node}'s line is elevated because it contains the keyword \texttt{alert}, while \texttt{boot\_init}'s lines are elevated because they are logged by a central subsystem that interfaces with unstandardized third-party components. For this incident, progressive onset tracking evaluates 7.9~million log lines across a 7-day window surrounding the upgrade. \systemname{} partitions this span into 2-hour tumbling windows to detect fault onset and construct the bipartite entity-event graph. Figure~\ref{fig:real_bipartite_masked} shows an excerpt of the maximal connected subgraph isolated by computing its transitive closure (Section~\ref{sec:design}).

End-to-end, \systemname{} processes this bundle---ingestion through precedence-forest construction---in under 48~minutes (Figure~\ref{fig:hardware_metrics_split}, labeled \textit{upgrade}), against the month engineers spent debugging it out-of-band.

In Figure~\ref{fig:real_bipartite_masked}, extracted high-entropy identifiers form entity nodes (top row), encompassing both runtime entity names (\texttt{boot\_init}, \texttt{config\_store}) and system entity names (\texttt{config\_store\_backup}, \texttt{data\_file\_store}). Because a log line's entity set defaults to its logging subsystem (Table~\ref{tab:constructs}), \texttt{boot\_init} is both the logging subsystem of the restore-sequence lines and an entity node in the graph. Their associated log statements form event nodes (bottom row), with undirected edges connecting each entity to its co-occurring events. Entities bind related events into connected components; for instance, \texttt{boot\_init} co-occurs across more than 570 events in the full graph, of which the figure shows five, spanning from the database connection failure and backup-restore sequence ($t_0 + 12\text{m}11\text{s}$ to $t_0 + 12\text{m}11.3\text{s}$) to the virtual switch restore attempt ($t_0 + 12\text{m}21\text{s}$).

\begin{figure*}[tb]
\centering
\small
\begin{tikzpicture}[node distance=0.45cm and 0.4cm,
  every node/.style={draw, rounded corners, align=center, font=\tiny}]

\node (title) [draw=none, font=\bfseries\scriptsize, anchor=west] at (-0.3,1.6) {Precedence Forest: 3 trees, 10 entities};

\node (up) [draw=none, dashed, font=\tiny, align=center] at (0,0.9) {$\vdots$\\ earlier ancestor\\ (hardware/BMC)};

\node (bi) [below=0.5cm of up, fill=gray!8] {\textbf{\texttt{boot\_init}}\\ $t_{\mathrm{beg}}{=}t0{+}12\text{m}11\text{s}$, $t_{\mathrm{end}}{=}t0{+}12\text{m}21\text{s}$};
\node (cs) [below left=0.6cm and -0.2cm of bi, fill=gray!8] {\textbf{\texttt{config\_store}}\\ $t0{+}12\text{m}11\text{s}$};
\node (vs) [below right=0.6cm and 1.6cm of bi, fill=gray!8] {\textbf{\texttt{virtual\_switch\_info}}\\ $t0{+}12\text{m}21\text{s}$};
\node (vl) [below=0.6cm of vs, fill=gray!8] {\textbf{\texttt{virtsw\_local\_binding}}\\{}[DEAD\_END]\\ $t0{+}12\text{m}21\text{s}$};

\node (ls) [below left=0.9cm and 2.3cm of cs, fill=gray!8, text width=1.5cm] {\textbf{\texttt{load\_store}}\\{}[DEAD\_END]\\ $t0{+}12\text{m}11.2\text{s}$};
\node (csb) [below=1.7cm of cs, fill=gray!8, text width=2.0cm] {\textbf{\texttt{config\_store\_backup}}\\{}[DEAD\_END]\\ $t0{+}12\text{m}11.2\text{s}$};
\node (dfs) [below right=0.9cm and 2.3cm of cs, fill=gray!8, text width=1.6cm] {\textbf{\texttt{data\_file\_store}}\\ $t0{+}12\text{m}11.2\text{s}$};
\node (dfb) [below=0.9cm of dfs, fill=gray!8, text width=1.75cm] {\textbf{\texttt{data\_file\_backup}}\\{}[DEAD\_END]\\ $t0{+}12\text{m}11.2\text{s}$};

\draw[->, thick] (cs) -- (bi);
\draw[->, thick] (vs) -- (bi);
\draw[->, thick] (vl) -- (vs);
\draw[->, thick] (ls) -- (cs);
\draw[->, thick] (csb) -- (cs);
\draw[->, thick] (dfs) -- (cs);
\draw[->, thick] (dfb) -- (dfs);
\draw[->, dashed] (bi) -- (up);

\node (av) [right=4.0cm of bi, very thick, dashed] {\textbf{\texttt{add\_virtual\_switch\_to\_host}}\\{}[ROOT\_CANDIDATE] [DEAD\_END]\\ $t_{\mathrm{beg}}{=}t0{+}0\text{s}$, $t_{\mathrm{end}}{=}t0{+}24\text{m}$};

\node (cp) [below=2.2cm of av, very thick, dashed] {\textbf{\texttt{config\_parse\_exception}}\\{}[ROOT\_CANDIDATE] [DEAD\_END]\\ $t0{-}37\text{m}$};

\node (moretrees) [below=0.6cm of cp, draw=none, dashed, font=\tiny, align=center] {$\cdots$ additional trees};

\end{tikzpicture}
\caption{\textbf{Incident Precedence Forest Excerpt.} Directed edges point from child entities to their nearest earlier co-occurring parent entities. Tree~A resolves 2 hops from \texttt{boot\_init} down to \texttt{virtsw\_local\_binding}, while Trees~B and C form isolated root candidates.}
\label{fig:real_forest_masked}
\end{figure*}
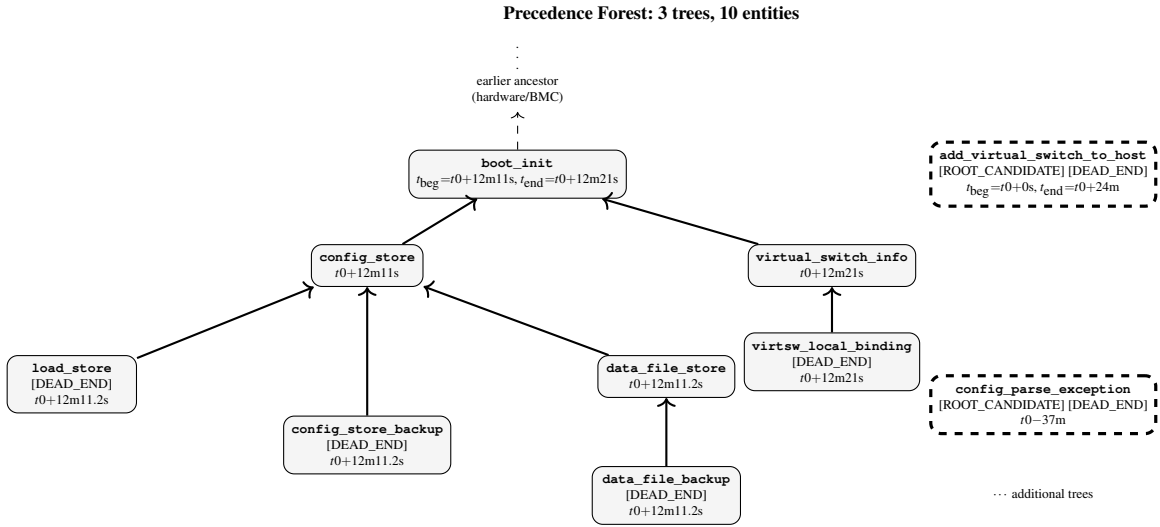

Linking each entity to its nearest-earlier co-occurring entity while pruning ubiquitous hub nodes transforms this bipartite graph into the precedence forest (Figure~\ref{fig:real_forest_masked}), partitioning the failure space into three distinct propagation trees. \textbf{Tree A (Connected Restore Branch)} is anchored at \texttt{boot\_init} (which links upward to an excluded hardware/BMC ancestor); this branch links the database failure to the backup-restore sequence (\texttt{load\_store}, \texttt{config\_store\_backup}, \texttt{data\_file\_store}, \texttt{data\_file\_backup}) and resolves two hops down to the failing switch entity (\texttt{virtsw\_local\_binding}). \textbf{Tree B (Isolated Retrying Root)} consists of \texttt{add\_virtual\_switch\_to\_host}, an isolated root candidate whose 13 retries share no entity tokens with Tree A, leaving it disconnected. \textbf{Tree C (Pre-existing Parsing Root)} is the separate, isolated root node \texttt{config\_parse\_exception} ($t_0 - 37\text{m}$). Each node in the forest encapsulates an impacted entity along with its associated co-occurring events.

While the bipartite graph in Figure~\ref{fig:real_bipartite_masked} captures how entities and events relate structurally, the precedence forest captures the temporal unfolding of failure propagation across impacted entities. Although the root cause of the incident itself lies outside the evaluated 7-day window, the forest establishes that within this window, the configuration parsing exception (\texttt{config\_parse\_exception}) strictly preceded all other faults in Figure~\ref{fig:real_forest_masked}. Meanwhile, failed retries by the host management daemon to attach the switch endpoint (\texttt{add\_virtual\_switch\_to\_host}) occurred concurrently with \texttt{boot\_init} faults. This temporal precedence enables operators to reconstruct the precise sequence of runtime and system entity faults during triage.

\noindent\textbf{LLM Reasoning over Topology.}
\label{sec:appendix_llm_transcript}
Because this experiment strictly withholds event text, the prompt frames the
input purely as structure: it explains that the data is a forest of
independent failure-propagation trees rather than a single chain, that each
entity is shown only by its first- and last-occurrence timestamps and its
precedence edges to other entities, and that certain entities are tagged as
likely entry points (no earlier identified cause within their own tree) or
terminal effects (no further downstream effect within their own tree). The
model is asked to trace forward from a tagged entry point along these edges
and propose a root cause and propagation path using only the graph's own
shape and entity naming---since no log content is available in this
ablation, the model cannot verify that its chosen entity's own line
describes an actual fault, a limitation the response below surfaces
directly.

Constructing the precedence forest reduces the diagnostic search space from 7.9~million lines over the 7-day period to 34,927 nodes, a $\sim 227\times$ reduction. To evaluate whether graph topology alone is sufficient for triage, we serialized a 1,600-node excerpt from this forest using only entity names, initial and final fault timestamps ($t_{\mathrm{beg}}, t_{\mathrm{end}}$), parent-child edges, and structural annotations (\texttt{[ROOT\_CANDIDATE]}, \texttt{[DEAD\_END]})---\textit{strictly omitting all raw log event text}---and queried Gemini for interpretation. The prompt yielded the following diagnostic response and self-reported caveats:

\begin{tcolorbox}[breakable, boxrule=0.4pt, arc=1pt, left=3pt, right=3pt, top=3pt, bottom=3pt, colback=gray!5, colframe=black!50]
\small\textit{``The \texttt{config\_parse\_exception} entity appears at $t_0-37\text{m}$, making it the earliest event in the timeline by a wide margin. This indicates a corrupted or malformed configuration file. This underlying failure likely caused the \texttt{add\_virtual\_switch\_to\_host} operation to hang or continuously fail starting at $t_0$ (lasting 24 minutes). The system or service was likely restarted to recover, evidenced by \texttt{boot\_init} at $t_0 + 12\text{m}11\text{s}$. During initialization, the system attempted to load the config and had to fall back to \texttt{config\_store\_backup} and \texttt{data\_file\_backup}, semantically corroborating that the primary configuration was broken by the earlier parse exception. \textbf{[Caveats]} Because these entities exist in separate trees without explicit co-occurrence edges, the causal links between the config exception, the hanging virtual switch operation, and the subsequent reboot are inferred solely from chronological sequencing and semantic naming. Furthermore, since \texttt{boot\_init} has an excluded hardware/BMC parent in the full graph, it is possible an underlying hardware or storage failure caused both the initial config file corruption (the parse exception) and a later forced hardware reboot, making the hardware fault the true unseen root cause.''}
\end{tcolorbox}

This response illustrates the power of bounding and ordering the log search space before incident triage. The precedence forest helps Gemini arrive at a hypothesis close to the actual cause: a configuration issue, without any domain-specific augmentation, despite the absence of event text. With domain expertise and event text, we expect human operators and downstream AIOps pipelines to improve over this response.

This example also highlights two gaps when applying LLM reasoning over the precedence forest. First, Gemini bridges disconnected trees into a single incident trajectory by leveraging temporal sequencing and semantic hints embedded within entity names---overriding the explicit structural boundaries maintained by \systemname{} due to a lack of co-occurring entities. Although ground truth from the incident confirms that all three trees indeed stem from the same cause (unpersisted configuration updates), this behavior reveals an LLM tendency to speculate cross-tree causality beyond explicit structural evidence. Second, because the latent root cause, a product defect, was unrecorded in the log bundle, the precedence forest captures only the subsequent failure propagation, which Gemini misattributes to hardware failures and file corruption due to the absence of event text. These gaps help identify the augmentations needed for a downstream AIOps pipeline: (1)~prompting must enforce explicit awareness of forest topology boundaries, and (2)~event text must be included to eliminate speculative hardware failure and file corruption hypotheses. We use the event text in our subsequent evaluations (\S\ref{sec:evaluation}).

%% file: sections/design.tex
\section{Design}
\label{sec:design}

In this section, we describe how the key concepts introduced in Section~\ref{sec:intro}—\textbf{severity elevation}, \textbf{progressive onset tracking}, and \textbf{hub-node pruning}—are combined to bound and order the diagnostic search space. \systemname{} is structured as a two-stage pipeline that transforms raw text logs into compact precedence forests. 

First, the \textit{ingestion stage} parses raw log lines into entity identifiers and log events, applying severity elevation to isolate fault candidates. Second, the \textit{graph construction stage} leverages progressive onset tracking and hub-node pruning to transform these fault candidates into compact precedence forests. Table~\ref{tab:constructs} summarizes the formal constructs, notation, and graph operations used across both stages.

\noindent\textbf{Log Ingestion.}
The ingestion stage transforms raw log lines $L$ into a filtered candidate fault set $E$ ($L \to E$). Using regex templates (\S\ref{sec:background}), raw lines sharing the same log pattern are collated ($\mathrm{CollateLines}(L)$) into parameterized events $e \in E$, inheriting the earliest line timestamp $t(e) = \min_{\ell \in e} t(\ell)$ and constituent entity sets $\text{Entities}(e) = \bigcup_{\ell \in e} \text{Entities}(\ell)$. To capture misclassified third-party errors (\S\ref{sec:intro}), \systemname{} applies severity elevation, promoting lines with embedded severity keywords or originating from central wrapper logs. The resulting elevated event stream $E = \{ e \in \mathrm{CollateLines}(L) \mid \mathrm{severity}(e) \ge \mathtt{[WARN]} \}$ serves as the primary input for graph construction.

\begin{table*}[t]
\centering
\small
\caption{\textbf{\systemname{} Constructs \& Notation.}}
\label{tab:constructs}
\begin{tabular}{@{}lp{0.7\linewidth}@{}}
\toprule
\textbf{Construct} & \textbf{Formal Definition \& Operations} \\ \midrule
\textbf{Raw Log Line} ($\ell \in L$) & Log entry containing timestamp $t(\ell)$, raw text, and logging subsystem $\text{Subsystem}(\ell)$. Entity set $\text{Entities}(\ell)$ is extracted via regular expressions, defaulting to $\{\text{Subsystem}(\ell)\}$. \\ \addlinespace
\textbf{Log Event} ($e \in E$) & Template-level aggregation $e = \text{CollateLines}(L)$ following severity elevation, with timestamp $t(e) = \min_{\ell \in e} t(\ell)$. Establishes entity correlation ($u, v \in \text{Entities}(e)$) and event correlation ($\text{Entities}(e_1) \cap \text{Entities}(e_2) \neq \emptyset$). \\ \addlinespace
\textbf{Log Window} ($w_i \subset L$) & Chronological $T_w = 2\text{h}$ slice used during progressive onset tracking across log pool $L$. \\ \addlinespace
\textbf{Maximal Failure Subgraph} ($F^* \subseteq E$) & Maximal connected component $F^* = \mathrm{argmax}_{C \in \mathcal{C}} \, \left| \bigcup_{e \in C} \text{Entities}(e) \right|$ selected from connected components $\mathcal{C}$ formed via transitive closure over the bipartite entity-event graph within correlation bound $|t(e_1) - t(e_2)| \le T_c = 2\text{h}$. Footprint: $\text{Subsystems}(F^*) = \{ \text{Subsystem}(e) \mid e \in F^* \}$. \\ \addlinespace
\textbf{Precedence Forest} ($\sigma$) & Graph $\sigma = (V, E_{\pi})$ constructed over maximal component $F^*$ via hub-node pruning:
\begin{itemize}[leftmargin=*, nosep]
    \item \textit{Hub-Node Blocklist ($G$):} Ubiquitous hub entity identifiers (e.g., global session identifiers or cluster-wide task handles).
    \item \textit{Impacted Entity Nodes ($V$):} Entities $\big(\bigcup_{e \in F^*} \text{Entities}(e)\big) \setminus G$, bound to $[t_{\mathrm{beg}}(v), t_{\mathrm{end}}(v)]$.
    \item \textit{Precedence Edges ($E_{\pi}$):} Directed pairs $(u^*, v)$ linking entity $v$ to its nearest temporally preceding co-occurring entity $u^*$.
\end{itemize} \\ \bottomrule
\end{tabular}
\end{table*}

\subsection{Graph Construction}
\label{sec:graph_construction}

The graph construction stage transforms the elevated fault set $E$ into compact precedence forests $\sigma$ ($E \to F^* \to \sigma$). Progressive onset tracking (Algorithm~\ref{alg:onset_tracking}) partitions $E$ into uniform temporal windows and detects fault propagation onset at the first window where correlated faults span two or more subsystems. From this onset window, it correlates faults across consecutive windows to construct a bipartite entity-event graph, computing its transitive closure to track the maximal connected component $F^* \subseteq E$ representing the primary failure propagation \textit{cascade} (Algorithm~\ref{alg:unionfind}). When a subsequent window adds no new faults to this component, graph expansion terminates. At both the initial onset and upon termination, \systemname{} converts $F^*$ into a precedence forest $\sigma$ via hub-node pruning and temporal edge assignment between impacted entities (Algorithm~\ref{alg:precforest}).

\noindent\textbf{Progressive Onset Tracking.}
To detect propagating failures without assuming a known outage timestamp, \systemname{} discretizes the timeline into uniform 2-hour windows ($W = (w_0, \dots, w_{N-1})$, $T_w = 2\text{h}$). These windows serve as input to progressive onset tracking (Algorithm~\ref{alg:onset_tracking}). Across these windows, \systemname{} tracks failure propagation \textit{onset}, \textit{expansion}, and \textit{termination} to output an ordered sequence of precedence forests $\mathcal{S} = \langle \sigma_1, \sigma_2, \dots \rangle$, where each forest $\sigma \in \mathcal{S}$ captures an isolated failure propagation cascade.

\noindent\textit{Onset.} For every window scanned, progressive onset tracking constructs a bipartite entity-event graph and computes its transitive closure to evaluate candidate failure components (Line~3). An onset is flagged when the subsystem footprint of the maximal connected component $F^*$ spans two or more distinct subsystems ($|\text{Subsystems}(F^*)| \ge K_{\mathrm{comp}} = 2$, Line~3), signaling that correlated faults are propagating across subsystem boundaries—i.e., a cascade is forming.

\noindent\textit{Expansion.} Upon identifying an onset window $w_{\text{onset}}$, the algorithm initiates an expanding window walk across up to $K_{\mathrm{walk}} = 15$ consecutive windows (Lines~7--19). By accumulating log lines $R = \bigcup w_i$, \systemname{} continuously recomputes maximal component $F_s^*$ to capture failure cascades that cross window boundaries.

\noindent\textit{Termination.} The expanding walk terminates either by naturally converging—when the failure cascade stops expanding ($F_s^* = F_{\text{prev}}^*$, Line~13)—or forcibly upon reaching the walk cap $K_{\mathrm{walk}} = 15$ without convergence (Line~20).

\noindent\textit{Forest Sequence.} At both initial onset ($s = 0$) and upon convergence ($F_s^* = F_{\text{prev}}^*$), the algorithm invokes \textsc{ConstructForest} to transform the maximal failure subgraph $F^*$ into a precedence forest $\sigma$ (Lines~13--14), enforcing temporal precedence across impacted entities. Capturing forests at both onset and upon convergence preserves both the initial failure cascade right after it impacts multiple subsystems, and the final cascade after convergence (or walk termination). If the walk cap $K_{\mathrm{walk}}$ is reached without converging, a fallback forest is emitted for the accumulated window slice (Line~21). The search index then advances past the processed cascade ($\text{idx} \gets i_{\text{last}} + 1$, Line~23) to resume scanning. Once the primary loop completes, any trailing unexamined windows ($\text{last\_covered} < N - 1$) undergo a final fallback scan (Lines~25--28), returning the complete sequence of precedence forests $\mathcal{S}$.

\begin{algorithm*}[t]
\caption{Progressive Onset Tracking}
\label{alg:onset_tracking}
\begin{algorithmic}[1]
\Require Log windows $W = (w_0, \dots, w_{N-1})$, correlation bound $T_c$, walk cap $K_{\mathrm{walk}} = 15$, onset threshold $K_{\mathrm{comp}} = 2$
\Ensure Sequence of precedence forests $\mathcal{S}$
\State $\mathcal{S} \gets \langle \rangle$, \quad $\text{idx} \gets 0$, \quad $\text{last\_covered} \gets -1$
\While{$\text{idx} < N$}
    \State $\mathcal{I} \gets \left\{ j \in [\text{idx}, N-1] \;\middle|\; \left| \text{Subsystems}\big(\textsc{DetectSubgraphs}(\text{CollateLines}(w_j), T_c)\big) \right| \ge K_{\mathrm{comp}} \right\}$
    \If{$\mathcal{I} = \emptyset$} \textbf{break} \EndIf
    \State $\text{onset} \gets \min(\mathcal{I})$, \quad $R \gets \emptyset$, \quad $F_{\text{prev}}^* \gets \emptyset$, \quad $i_{\text{last}} \gets \text{onset} - 1$, \quad $\text{terminated} \gets \text{false}$
    \For{$s \gets 0$ \To $K_{\mathrm{walk}} - 1$}
        \If{$\text{onset} + s \ge N$} \textbf{break} \EndIf
        \State $i_{\text{last}} \gets \text{onset} + s$ \Comment{Track window index currently processed}
        \State $R \gets R \cup w_{i_{\text{last}}}$
        \State $F_s^* \gets \textsc{DetectSubgraphs}(\text{CollateLines}(R), T_c)$
        \If{$s = 0 \lor F_s^* = F_{\text{prev}}^*$} \Comment{Initial onset or natural convergence}
            \State $\mathcal{S} \gets \mathcal{S} \mathbin{\|} \langle \textsc{ConstructForest}(F_s^*, R) \rangle$
            \If{$s > 0$} $\text{terminated} \gets \text{true}$; \textbf{break} \EndIf
        \EndIf
        \State $F_{\text{prev}}^* \gets F_s^*$
    \EndFor
    \If{\AlgNot $\text{terminated} \land s > 1$} \Comment{Walk cap reached without natural convergence}
        \State $\mathcal{S} \gets \mathcal{S} \mathbin{\|} \langle \textsc{ConstructForest}(F_{\text{prev}}^*, R) \rangle$ \Comment{Emits $F_{\text{prev}}^*$ for last processed window}
    \EndIf
    \State $\text{last\_covered} \gets \max(\text{last\_covered}, i_{\text{last}})$, \quad $\text{idx} \gets i_{\text{last}} + 1$
\EndWhile
\If{$\text{last\_covered} < N - 1$} \Comment{Fallback scan for trailing window}
    \State $F_{\text{final}}^* \gets \textsc{DetectSubgraphs}(\text{CollateLines}(w_{N-1}), T_c)$
    \State $\mathcal{S} \gets \mathcal{S} \mathbin{\|} \langle \textsc{ConstructForest}(F_{\text{final}}^*, w_{N-1}) \rangle$
\EndIf
\State \Return $\mathcal{S}$
\end{algorithmic}
\end{algorithm*}

\noindent\textbf{Bounded Bipartite Subgraph Detection.}
\textsc{DetectSubgraphs} correlates input faults $E$ within a temporal correlation bound $T_c = 2\text{h}$ to construct a bipartite entity-event graph. Within this graph, it isolates failure propagation cascades by computing its transitive closure, returning the maximal failure subgraph $F^* \subseteq E$ representing the primary failure cascade (Algorithm~\ref{alg:unionfind}).

\begin{algorithm}[t]
\caption{Bounded Bipartite Subgraph Detection}
\label{alg:unionfind}
\begin{algorithmic}[1]
\Function{DetectSubgraphs}{$E, T_c$}
    \State $\mathrm{UF} \gets \text{Initialize UnionFind over log events } E$
    \For{each entity $u \in \bigcup_{e \in E} \text{Entities}(e)$}
        \State Let $L_u = (e_1, e_2, \dots, e_m)$ be log events in $E$ containing $u$, ordered by $t(e)$
        \For{$i \gets 1$ \To $m - 1$}
            \If{$t(e_{i+1}) - t(e_i) \le T_c$}
                \State $\mathrm{UF.union}(e_i, e_{i+1})$
            \EndIf
        \EndFor
    \EndFor
    \State $\mathcal{C} \gets \text{ConnectedComponents}(\mathrm{UF})$
    \State $F^* \gets \mathrm{argmax}_{C \in \mathcal{C}} \, \left| \bigcup_{e \in C} \text{Entities}(e) \right|$ \Comment{Maximal bounded bipartite subgraph}
    \State \Return $F^*$
\EndFunction
\end{algorithmic}
\end{algorithm}

To compute transitive closure across shared entities efficiently, the algorithm initializes a Union-Find structure over $E$ (Line~2), whose amortized near-linear complexity is a classical result~\cite{tarjan1975unionfind}. For each unique entity $u$, \systemname{} constructs a chronologically ordered sequence of its associated faults $L_u = (e_1, e_2, \dots, e_m)$ (Lines~3--4). To bound search-space growth temporally and avoid correlating faults far apart in time, \systemname{} merges adjacent faults $e_i$ and $e_{i+1}$ via \textsf{union} only when their temporal separation falls within the correlation bound $t(e_{i+1}) - t(e_i) \le T_c$ (Lines~5--8). Finally, the algorithm extracts all connected components $\mathcal{C}$ (Line~11) and selects $F^* = \mathrm{argmax}_{C \in \mathcal{C}} \, \left| \bigcup_{e \in C} \text{Entities}(e) \right|$ as the maximal connected component spanning the most impacted entities (Line~12), returning $F^*$ as the maximal failure subgraph for precedence forest construction (Line~13).

\noindent\textbf{Precedence Forest Construction.}
\textsc{ConstructForest} accepts the maximal failure subgraph $F^*$ and accumulated log lines $R$, returning a precedence forest $\sigma = (V, E_{\pi})$ that establishes the chronological fault propagation topology across impacted entities (Algorithm~\ref{alg:precforest}).

To prevent ubiquitous infrastructural handles from acting as false root hubs (hub-node pruning), the algorithm first filters entities in $F^*$ against blocklist $G$ to isolate the impacted node set $V$ (Line~2). Next, \systemname{} records the first and last appearance of each entity $v \in V$ across log lines $R$ as its active temporal boundaries $[t_{\mathrm{beg}}(v), t_{\mathrm{end}}(v)]$ (Lines~3--6), defining its active event span for triage (\S\ref{sec:implementation}). Sorting $V$ in ascending order by earliest appearance—breaking timestamp ties deterministically by entity ID $v$, i.e., $(t_{\mathrm{beg}}(v), v)$—establishes a global rank $\text{order}(v)$, capturing each entity's earliest fault since onset (Line~7).

\begin{algorithm}[t]
\caption{Precedence Forest Construction}
\label{alg:precforest}
\begin{algorithmic}[1]
\Function{ConstructForest}{$F^*, R$}
    \State $V \gets \left( \bigcup_{e \in F^*} \text{Entities}(e) \right) \setminus G$ \Comment{Hub-node pruning}
    \For{each entity $v \in V$} \Comment{Phase 1: Record active temporal boundaries}
        \State $t_{\mathrm{beg}}(v) \gets \min \{ t(\ell) \mid \ell \in R \land v \in \text{Entities}(\ell) \}$
        \State $t_{\mathrm{end}}(v) \gets \max \{ t(\ell) \mid \ell \in R \land v \in \text{Entities}(\ell) \}$
    \EndFor
    \State $\text{order}(v) \gets$ rank of $v$ when $V$ is sorted ascending by $(t_{\mathrm{beg}}(v), v)$ \Comment{Phase 2: Global order}
    \State $E_{\pi} \gets \emptyset$
    \For{each entity $v \in V$} \Comment{Phase 3: Parent pointer assignment}
        \State $C(v) \gets \{ u \in V \setminus \{v\} \mid \exists e \in F^* \text{ s.t. } \{u, v\} \subseteq \text{Entities}(e) \}$ \Comment{Co-occurring entities}
        \State $P(v) \gets \{ u \in C(v) \mid \text{order}(u) < \text{order}(v) \}$ \Comment{Temporally preceding candidates}
        \If{$P(v) \neq \emptyset$}
            \State $u^* \gets \mathrm{argmax}_{u \in P(v)} \, \text{order}(u)$ \Comment{Nearest-earlier co-occurring parent entity}
            \State $E_{\pi} \gets E_{\pi} \cup \{ (u^*, v) \}$
        \EndIf
    \EndFor
    \State \Return $\sigma = (V, E_{\pi})$ \Comment{Precedence forest}
\EndFunction
\end{algorithmic}
\end{algorithm}

Finally, the algorithm constructs parent-child precedence edges $E_{\pi}$ (Line~8). For each entity $v \in V$, \systemname{} identifies co-occurring entities in $F^*$ (Line~10) and filters for temporally preceding candidates $P(v)$ where $\text{order}(u) < \text{order}(v)$ (Line~11). If preceding candidates exist, a directed precedence edge $(u^*, v)$ is drawn from the single nearest prior co-occurring entity $u^* = \mathrm{argmax}_{u \in P(v)} \, \text{order}(u)$ (Lines~12--14). This yields a directed acyclic forest $\sigma = (V, E_{\pi})$ (Line~17), producing a compact, ordered diagnostic artifact suited for direct human triage or downstream AIOps synthesis.

%% file: sections/impl.tex
\section{Implementation}
\label{sec:implementation}

\systemname{} is implemented as a CPU-only pipeline in approximately 3,150 lines of Python 3.10+. It leverages Python's \texttt{concurrent.futures} thread pools and pre-compiled regular expressions for parallel log ingestion, utilizing SQLite (v3.35+) to persist parsed event templates and extracted entities for retrieval by progressive onset tracking.

\noindent\textbf{Vendor-Specific Log Parsing.}
\systemname{} employs lightweight, source-specific integration modules for both enterprise environments and open-source benchmarks (OpenStack~\cite{openstack}, TrainTicket~\cite{trainticket}). Because enterprise log schemas remain stable in production, each log source requires only a one-time integration module with four components: (1)~a pre-compiled regex cascade extracting the primary 4-field tuple ($\text{timestamp}, \text{subsystem}, \text{severity}, \text{text}$); (2)~a semantic severity elevation rule set targeting embedded error keywords and central orchestrator subsystems; (3)~entity extraction patterns (e.g., \texttt{config\_store} UUIDs in virtual infrastructure, TraceIDs/SpanIDs in TrainTicket); and (4)~a hub-node blocklist $G$ specifying ubiquitous infrastructure handles.

\noindent\textbf{Intermediate Persistence \& Storage.}
Parsed events and entities from log ingestion are streamed into per-subsystem SQLite shards in write-batch flushes of 5,000 items. This sharded write bounds ingestion memory to a flat RSS footprint regardless of raw log scale $N$. The normalized event templates $E$ with \verb+<ENTITY>+ placeholders are computed per log line and stored in the database. Indexes on severity and timestamps enable fast window queries over faults ($N'$) during progressive onset tracking, while raw log files remain on disk for lazy parameter resolution. The constructed precedence forests $\sigma$—comprising entities, event templates, and precedence edges—are persisted in the database layer for triage.

%% file: sections/eval.tex
\section{Evaluation}
\label{sec:evaluation}
Our work aims to address a key pain point of enterprise operators facing black-box outages: the lack of domain artifacts to diagnose incidents. This section evaluates how effectively \systemname{} closes this gap by utilizing strictly raw logs across 25 enterprise outages and 12 open-source software incidents. All evaluations are conducted on a VM with $8$~vCPUs, $256\text{~GB}$ RAM, running Ubuntu 22.04 LTS. LLM interpretability is evaluated using Google's \texttt{gemini-3.1-pro-preview} model via the Gemini Enterprise API endpoint on Vertex AI (temperature $T=0.1$, default decoding parameters).

Our evaluation answers five core questions: (1)~\textbf{Search-Space Bounding:} How effectively does \systemname{} isolate ground-truth root causes while eliminating background noise (\S\ref{sec:recall})? (2)~\textbf{Diagnostic Utility:} Can precedence forests enable zero-shot LLM root-cause identification without domain artifacts (\S\ref{sec:interpret})? (3)~\textbf{Early Detection:} Does progressive onset tracking provide actionable lead time and alert reduction (\S\ref{sec:onset})? (4)~\textbf{Sensitivity \& Ablation:} How do \systemname{}'s parameter settings and individual design decisions impact recall, compression, and diagnostic accuracy (\S\ref{sec:sensitivity})? (5)~\textbf{System Overhead:} What compute and memory runtime costs does \systemname{} impose (\S\ref{sec:overheads})?

\subsection{Forest Recall and Accuracy}
\label{sec:recall}

\begin{table}[t]
\centering
\scriptsize
\caption{\textbf{Top-$k$ ($k \in \{1,10\}$) reachability and recall for graph baselines vs. \systemname{} on enterprise incidents.} Closeness, Betweenness, PPR, and earliest-temporal order are computed on the identical 7-day/$T_c=2$h (and optionally hub-node pruned) graph \systemname{} uses across $N=16$ tractable enterprise bundles. \systemname{} is reported for both populations: restricted to $N=16$ for a direct baseline comparison, and across the full $N=25$ enterprise population.}
\label{tab:stage1_full168h_final}
\begin{tabular}{@{}l rrrr@{}}
\toprule
\textbf{Method} & \textbf{Rank-1} & \textbf{Top-10} & \textbf{Recall@All} & \textbf{Recall@1600} \\ \midrule
Closeness & 0.0\% & 12.5\% & 0.60 & 0.33 \\
Closeness (pruned) & 0.0\% & 12.5\% & 0.58 & 0.31 \\
Betweenness & 0.0\% & 18.8\% & 0.60 & 0.38 \\
Betweenness (pruned) & 6.2\% & 18.8\% & 0.58 & 0.36 \\
PPR~\cite{wu2020microrca,kim2013root} & 0.0\% & 31.2\% & 0.60 & 0.44 \\
PPR~\cite{wu2020microrca,kim2013root} (pruned) & 6.2\% & 31.2\% & 0.58 & 0.42 \\
Earliest-temporal order & 0.0\% & 6.2\% & 0.60 & 0.41 \\
Earliest-temporal order (pruned) & 0.0\% & 12.5\% & 0.58 & 0.38 \\ \midrule
\textbf{\systemname{} ($N=16$)} & \textbf{31.2\%} & \textbf{75.0\%} & \textbf{0.72} & \textbf{0.55} \\ \midrule
\textbf{\systemname{} ($N=25$)} & \textbf{20.0\%} & \textbf{60.0\%} & \textbf{0.76} & \textbf{0.46} \\ \bottomrule
\end{tabular}
\end{table}

\begin{table}[t]
\centering
\scriptsize
\caption{\textbf{Top-$k$ ($k \in \{1,10\}$) reachability and recall for graph baselines vs. \systemname{} on open-source software incidents.} Recall@1600 is identical to Recall@All ($1.00$), hence not shown. Hub-node pruning yields identical performance across all baselines and is omitted for brevity.}
\label{tab:stage1_oss_baselines}
\begin{tabular}{@{}l rrr@{}}
\toprule
\textbf{Method} & \textbf{Rank-1} & \textbf{Top-10} & \textbf{Recall@All} \\ \midrule
Closeness & 33.3\% & 33.3\% & 1.00 \\
Betweenness & 33.3\% & 33.3\% & 1.00 \\
PPR~\cite{wu2020microrca,kim2013root} & 33.3\% & 33.3\% & 1.00 \\
Earliest-temporal order & 33.3\% & 91.7\% & 1.00 \\ \midrule
\textbf{\systemname{}} & \textbf{83.3\%} & \textbf{100.0\%} & \textbf{1.00} \\ \bottomrule
\end{tabular}
\end{table}

To explain how a root cause leads to a symptom, the precedence forest must capture not only the cause and terminal symptom, but also \textit{intermediate faults}---mirroring a support engineer's diagnostic process. Failing to capture these intermediate faults degrades \textbf{recall}, while placing them in trees ranked late in the forest degrades \textbf{accuracy} by increasing the number of candidate trees an operator must inspect.

To evaluate recall and accuracy, we extract all impacted entities documented in vendor incident diagnoses and open-source post-mortems as ground truth (GT). \textbf{Recall} measures the fraction of GT terms (entity values and event metadata) retained in the extracted forest. Severity elevation isolates GT faults in 24 of 25 enterprise incidents, establishing an empirical recall ceiling of $0.96$ on enterprise telemetry (versus $1.00$ on open-source benchmarks); the one incident without isolated GT faults is a severity-elevation miss (1 of 25), and a further 3 of 25 incidents have zero recall because maximal cascade selection discards their root-cause subgraph (\S\ref{sec:limitations}), which together with partial retention in the remaining incidents yields the $0.76$ mean recall. Notably, all four of these zero-recall enterprise incidents (the severity-elevation miss and the three maximal-cascade-selection cases) stem from third-party components, demonstrating that unannotated third-party log streams present the steepest boundary for log-only correlation. To evaluate accuracy, we order trees in the precedence forest chronologically by their root node's earliest fault timestamp. We quantify accuracy via \textbf{Top-$k$ reachability}—the fraction of incidents where a GT-matching tree ranks within the first $k$ roots examined.

\noindent\textbf{Comparative Baseline Selection.} We compare \systemname{} against four canonical baselines that operate directly on raw telemetry without requiring pre-existing dependency maps or repeated observations (unlike Sherlock~\cite{bahl2007sherlock}): (1)~\textbf{Closeness Centrality} (shortest-path distance sums), (2)~\textbf{Personalized PageRank} (random walks with seed teleportation)~\cite{wu2020microrca,kim2013root}, (3)~\textbf{Betweenness Centrality} (shortest-path traversal density), and (4)~\textbf{Earliest-Temporal Order} (first-seen timestamp sorting). Proprietary frameworks (e.g., \texttt{RCACopilot}, \texttt{AIDA}) are excluded as they are closed-source. We similarly exclude log-anomaly baselines (DeepLog~\cite{du2017deeplog}, LogBERT~\cite{guo2021logbert}, LogAnomaly~\cite{meng2019loganomaly}): they mask entity parameters into wildcards and require healthy baselines this evaluation's incidents lack (see the LogBERT paragraph below). For strict experimental parity, all graph baselines execute on the exact same candidate graph constructed by \systemname{} following severity elevation.

Table~\ref{tab:stage1_full168h_final} summarizes performance across enterprise virtual infrastructure incidents. Due to $O(M^2)$ memory scaling (\S\ref{sec:background}), graph baselines run out of memory on 9 enterprise incidents, completing on only 16 tractable bundles. Across them, overall baseline recall (\textit{Recall@All}) is identical across methods within each variant ($0.60$ raw, $0.58$ hub-node pruned) because all four rank the same underlying entity set. This recall remains capped below \systemname{}'s ($0.72$ on $N=16$) because the baselines rank entities exclusively, ignoring ground-truth embedded in event pattern text. Baseline Top-10 reachability remains low ($\le 31.2\%$) because centrality metrics consistently favor high-volume, central subsystems—a topological bias that hub-node pruning only marginally mitigates. Earliest-temporal ordering similarly degrades when root-cause events occur later in the incident window.

By preserving its $O(M)$ bipartite structure, \systemname{} evaluates across all $N=25$ incidents. Because it ranks precedence trees rather than flat entity lists, any path from an impacted entity back to its root provides a complete fault propagation chain of entities and associated log events. Consequently, \systemname{} achieves $60.0\%$ Top-10 reachability and $0.76$ overall recall ($0.46$ prompt-capped at 1,600 nodes) across all 25 incidents—nearly double the accuracy of graph baselines, approaching the $0.96$ candidate fault isolation ceiling.

On open-source benchmarks (Table~\ref{tab:stage1_oss_baselines}), centrality baselines exhibit the same topological bias toward high-volume, central subsystems ($33.3\%$ Top-10 reachability). In contrast, earliest-temporal ordering reaches $91.7\%$ Top-10 reachability, reflecting stronger chronological signaling in these workloads. \systemname{} enforces temporal precedence over the bipartite entity-event graph without incurring centrality bias, achieving $83.3\%$ Rank-1 and $100.0\%$ Top-10 reachability.

\noindent\textbf{LogBERT: Healthy Baseline Requirement.}
To empirically ground the exclusion of stationary healthy-baseline methods, we evaluated LogBERT~\cite{guo2021logbert} across four datasets spanning from its intended operating regime to our target environment. To mirror live incident response, LogBERT was trained without synthetic baselines or dataset balancing. On \textbf{HDFS}~\cite{xu2009detecting}, which provides pre-partitioned block identifiers and a clean healthy-anomalous split, LogBERT achieves $F1=0.69$ ($P=0.96$, $R=0.54$), somewhat below the published results~\cite{guo2021logbert}. We exclude HDFS from \systemname{}'s primary evaluation because its single-subsystem failure model does not exercise cross-subsystem correlation. Outside this controlled setting, LogBERT fails to generalize. On \textbf{OpenStack}, severe class imbalance ($0.1\%$ anomalies) causes LogBERT to converge to the majority normal class ($99.9\%$ accuracy, $F1=0.00$). On \textbf{TrainTicket}, post-mortem log snapshots lack the pre-incident steady state required to learn a decision boundary, leading LogBERT to label all 3,412 evaluated sequences as anomalous ($F1=0.00$). On real enterprise field incidents, including the upgrade incident (\S\ref{sec:case_study}), LogBERT fails across every incident tested. This progression (strong performance on the lone benchmark matching its baseline assumption, complete failure on every dataset violating it) is the empirical counterpart to this section's argument: the assumption itself, not implementation quality, breaks in enterprise environments.

\subsection{Interpretability of Precedence Forests}
\label{sec:interpret}

To evaluate whether precedence forests enable domain-independent root-cause diagnosis, we measure zero-shot \emph{Diagnosis Quality (DQ)} and response consistency. A response earns a \emph{Full Diagnosis} if the LLM correctly identifies ground-truth (GT) entities and the root GT subsystem, and a \emph{Partial Diagnosis} if it identifies the root GT subsystem even when specific GT entities are omitted or truncated; otherwise it is a \emph{Miss}.

We evaluate DQ by prompting Gemini using the top 1,600 forest nodes. This 1,600-node prompt budget retains $0.46$ enterprise recall and $1.00$ open-source recall within the prompt itself (Tables~\ref{tab:stage1_full168h_final}--\ref{tab:stage1_oss_baselines}), imposing a hard ceiling on Full Diagnoses because the model can only quote entities present in its prompt context. To evaluate accuracy beyond this budget constraint, we apply a post-call \emph{subsystem-impact annotation}: \systemname{} walks each candidate's descendant subtree in the complete, uncapped precedence forest stored in the database and appends its real subsystem footprint to the output. A response is credited as a Partial Diagnosis if this augmented, database-verified footprint touches the GT subsystem, even if the specific GT entity was truncated from the 1,600-node prompt budget.

Unlike topology-only ablations (\S\ref{sec:appendix_llm_transcript}), every prompt in this evaluation includes each entity's first and last fault in the forest alongside its structural position as a root or leaf node (see \texttt{[ROOT\_CANDIDATE]} and \texttt{[DEAD\_END]} tags in Figure~\ref{fig:real_forest_masked}). The prompt frames the input as a forest of independent trees and leverages the model for semantic reasoning by instructing it to trace forward along each tree until reaching an explicit operational fault—regardless of whether that entity is temporally earliest—to prevent false attribution to benign health checks or initialization events. The prompt enforces a three-step reasoning chain:
\begin{enumerate}[leftmargin=16pt, topsep=2pt, itemsep=1pt]
  \item \textbf{Intra-Tree Propagation:} Connect each independent tree from its entry point to its terminal effect, explaining the step-by-step propagation between them.
  \item \textbf{Cross-Tree Semantic Linking:} Identify semantic links between trees to capture causal ties where entity co-occurrence was not logged.
  \item \textbf{Systemic Cause Synthesis:} Scan across all trees for recurring subsystems or concepts that explain multiple independent symptoms.
\end{enumerate}
Rather than collapsing this reasoning into a single forced answer, the model reports every credible candidate side by side with supporting evidence and confidence scores.

\begin{table}[t]
\centering
\scriptsize
\caption{\textbf{Diagnosis Quality (DQ) across enterprise and open-source software incidents.} Evaluated using Gemini on raw warning+ log text versus \systemname{}'s precedence forest capped at 1,600 nodes. Full Diagnosis indicates the response identified ground-truth (GT) entities together with the GT subsystem. Partial Diagnosis indicates that while the specific GT entities were omitted or truncated, the correct GT subsystem was identified.}
\label{tab:llm_diagnosis_quality}
\begin{tabular}{@{}l cc@{}}
\toprule
\textbf{Method} & \textbf{Full Diagnosis} & \textbf{Full + Partial} \\ \midrule
\multicolumn{3}{@{}l}{\textit{Enterprise Incidents}} \\ 
Gemini (Baseline) & 4 / 25 & 9 / 25 \\
\textbf{\systemname{} + Gemini} & \textbf{10 / 25} & \textbf{20 / 25} \\ \midrule
\multicolumn{3}{@{}l}{\textit{Open-Source Software Logs}} \\
Gemini (Baseline) & 9 / 12 & 10 / 12 \\
\textbf{\systemname{} + Gemini} & \textbf{10 / 12} & \textbf{12 / 12} \\ \bottomrule
\end{tabular}
\end{table}

\noindent\textbf{Results \& Diagnostic Accuracy.}
As Table~\ref{tab:llm_diagnosis_quality} shows, Gemini prompted with raw enterprise warning+ logs achieves Full Diagnoses in only 4/25 incidents (9/25 with Partial Diagnoses). Because 7-day raw bundles (median 55 GB) exceed context limits, this baseline evaluates a truncated stream of trailing warning+ events, meaning some baseline misses reflect context truncation rather than reasoning failure. Precedence forests increase Full Diagnoses to 10/25—tracking the $0.46$ capped-recall ceiling that governs exact entity matches—while post-call subsystem annotations add 10 Partial Diagnoses, reaching 20/25 (80\%) overall diagnostic accuracy. On open-source benchmarks, precedence forests raise overall accuracy from 10/12 to 12/12 (100\%). This gain is notable because raw open-source log streams already fit within Gemini's context window, proving structural distillation adds diagnostic value beyond simple token compression; the remaining gap on enterprise incidents also reflects proprietary vendor binaries omitting intermediate state that open-source stack traces expose directly. Subsystem ownership further explains this diagnostic boundary: \systemname{} achieves $100\%$ diagnostic accuracy ($9/9$ Full + Partial) on incidents originating within the core infrastructure software, whereas accuracy on emergent third-party component failures is $69\%$ ($11/16$). Crucially, all $5$ enterprise Misses occur on third-party software, where proprietary vendor binaries omit intermediate runtime state.

\noindent\textbf{Response Consistency \& Repeatability.}
Beyond single-execution accuracy, precedence forests significantly improve diagnostic repeatability across independent runs. Evaluating three independent trials per method on the matched population of 20 enterprise incidents (60 trials per condition), precedence forests sustain at least a Partial Diagnosis across all three trials in 18 of 20 incidents (90\%), compared to only 7 of 20 (35\%) for raw-log prompting. Across all 60 trial executions, precedence forests deliver at least a Partial Diagnosis in 95.0\% of runs (57/60)—with 46.7\% achieving a Full Diagnosis (28/60)—whereas raw-log prompting succeeds in only 35.0\% of runs (21/60) and reaches a Full Diagnosis in 13.3\% (8/60). While raw logs exhibit high nominal score agreement (17/20), this is driven by consistent failure (scoring zero across all three trials in 13 incidents); in contrast, precedence forests provide high-yield, repeatable diagnostic signal without trial-to-trial degradation.

\begin{table}[t]
\centering
\scriptsize
\setlength{\tabcolsep}{3pt}
\caption{\textbf{Alert lead times, volume, and search space reduction compared to enterprise reality (last 7 days of logs).} Lead times are evaluated on the 20 incidents with a Full or Partial Diagnosis, reported as the earliest checkpoint at which a real LLM call on the precedence forest achieves its best diagnosis quality for that incident (\S\ref{sec:appendix_casestudies}). Enterprise alert volume reflects $N=22$ bundles that recorded alerts shown to the administrator. \systemname{} metrics span all $N=25$ enterprise incidents.}
\label{tab:onset_operational_yield}
\begin{tabular}{@{}p{2.2cm}p{1.7cm}p{2.1cm}p{1.4cm}@{}}
\toprule
\textbf{Metric} & \textbf{Enterprise} & \textbf{\systemname{}} & \textbf{Impact} \\ \midrule
\textbf{Lead Time} & Reactive (0h) & Median 16h\newline(Mean 25.1h) & +0h to +144h \\ \addlinespace
\textbf{Alert Volume} & Median 1,658 & Median 12 & $124\times$ median per-incident reduction \\ \addlinespace
\textbf{Search Space} & Median 55~GB & Full: 133.8~MB\newline Forest@1600: 1.4~MB & $99.8\%$ reduction \\ \bottomrule
\end{tabular}
\end{table}

\subsection{Reduced and Early Alerting}
\label{sec:onset}
Alert floods slow down incident detection and response by overwhelming enterprise administrators, allowing developing issues to go unnoticed for days before an outage. To be effective, progressive onset tracking must detect failures early \textit{and} provide fewer, more meaningful alerts that connect system faults while bounding the diagnostic search space.

Each precedence forest constructed during progressive onset tracking acts as a discrete alert—generating two forests per walk (one at onset and another upon convergence). To measure detection earliness, we evaluate the earliest temporal checkpoint at which a precedence forest enables a real LLM call to reach its highest DQ for that incident. To evaluate alert reduction, we compare raw administrator alert volume over a 7-day window against total generated forests. Finally, to establish search-space reduction, we compare total 7-day log event volume against the event count in the final, maximal precedence forest. 

Table~\ref{tab:onset_operational_yield} summarizes our findings across the 25 enterprise incidents. For the 20 incidents with a Full or Partial Diagnosis (Table~\ref{tab:llm_diagnosis_quality}), \systemname{} achieves a median diagnosis-verified lead time of 16~hours (mean 25.1h) before the reported incident timestamp ($t_0$). Crucially, this lead time is verified empirically by evaluating LLM DQ (\S\ref{sec:interpret}) across every forest alert generated during progressive onset tracking (\S\ref{sec:appendix_casestudies}), proving that enterprise failures incubate long before escalation and can be correctly diagnosed hours in advance. Across 22 incidents with available alert telemetry, enterprise environments logged a median of 1,658 raw alerts in their final 7 days. In contrast, \systemname{} consolidates these streams into a median of 12 precedence forests over the same period (a median per-incident reduction of $124\times$), structuring individual alerts into coherent failure cascades. Finally, raw log bundles (median 55~GB, $\sim$2.5~million entries) are compressed into precedence forests with a median size of 133.8~MB (1.4~MB when capped at 1,600 nodes for LLM context), delivering a median search-space reduction of 99.8\%. The next subsection details five representative incidents, comparing precedence forests against a raw operator alert stream at the same lead-time checkpoint.

\begin{table*}[tb]
\centering
\small
\caption{\textbf{Early-detection case studies.} Lead time denotes hours before the reported incident timestamp ($t_0$) at which \systemname{}'s precedence forest isolates the fault signature to reach its earliest maximum-DQ checkpoint (\S\ref{sec:appendix_casestudies}), with raw alerts evaluated at the identical checkpoint. \emph{Miss} indicates failure to surface the ground-truth cause; \emph{false narrative} denotes an active misdirection where raw-alert prompting constructs an incorrect causal story from coincidental background alerts.}
\label{tab:case_study_summary}
\begin{tabular}{@{}p{6.4cm}cll@{}}
\toprule
\textbf{Incident} & \textbf{Lead Time} & \textbf{Forest (at lead time)} & \textbf{Raw alerts (at lead time)} \\ \midrule
Latent config-store persistence failure during hypervisor upgrade (\S\ref{sec:case_study}) & 18h & \textbf{Full} & Miss (false narrative) \\ \addlinespace
Intermittent storage adapter disconnections and target path flapping & 26h & \textbf{Full} & Partial \\ \addlinespace
Recurring fabric driver memory allocation failure and command queue exhaustion & 144h & Partial & Miss \\ \addlinespace
Storage-volume quiesce-timeout notification registration fault & 86h & Partial & Miss \\ \addlinespace
Session-scheduler configuration fault masked by coincidental network adapter flapping & 18h & Partial & Miss (false narrative) \\ \bottomrule
\end{tabular}
\end{table*}

\subsection{Case Studies: Early Detection}
\label{sec:appendix_casestudies}

This subsection grounds our 16-hour median diagnosis-verified lead time in five representative enterprise incidents, evaluating DQ at each incident's earliest maximum-DQ checkpoint across two representations: (1)~\systemname{}'s precedence forest, evaluated using the prompting and post-call subsystem-impact annotation methodology established in \S\ref{sec:interpret}, and (2)~the raw stream of log alerts an enterprise console would surface over the same window without \systemname{} (``raw alerts''). The earliest maximum-DQ checkpoint is identified by executing an LLM call (\texttt{gemini-3.1-pro-preview}) at every forest alert generated during progressive onset tracking, with raw alerts evaluated at the identical checkpoint to isolate the diagnostic value of precedence forests.

At their respective lead-time checkpoints, precedence forests enable a Full Diagnosis in two of five incidents (Cases 1--2) and a Partial Diagnosis in the remaining three (Cases 3--5). In contrast, raw alerts achieve a Partial Diagnosis in only one incident (Case 2) and fail in the other four—twice constructing confident, false narratives from coincidental background events (Cases 1 and 5). Table~\ref{tab:case_study_summary} summarizes these diagnostic outcomes at each incident's lead-time checkpoint.

\paragraph{Case 1: Latent config-store persistence failure during hypervisor upgrade.}
In this incident, the outage manifested at $t_0$ as VM migration failures following a hypervisor upgrade. The root cause was a latent product defect that prevented configuration updates from being persisted to disk, causing \texttt{config\_store} to restore a stale snapshot upon reboot (\S\ref{sec:case_study}). Prior to $t_0$, the incubating fault emitted background log alerts. At $t_0-80\text{h}$ and $t_0-56\text{h}$, the LLM reading raw alerts identified only routine automated management logins, correctly classifying them as benign. In contrast, precedence forests identified the impacted subsystems as early as $t_0-80\text{h}$, achieving a Partial Diagnosis. At $t_0-32\text{h}$, the raw-alert diagnosis shifted to a false narrative, attributing the alerts to an out-of-band hardware-management daemon and network interface flapping. At $t_0-18\text{h}$ (the earliest maximum-DQ checkpoint), \systemname{}'s precedence forest isolated the fault's signature to \texttt{config\_store}. It surfaced \texttt{config\_store}'s read/parse errors, a virtual switch port state-save failure, and a VM disk-cache failure to achieve a Full Diagnosis 18 hours prior to the reboot. The raw-alert diagnosis at $t_0-18\text{h}$ remained stuck on the hardware false narrative. At $t_0$, the outage occurred and the precedence forest maintained its Full Diagnosis anchored on \texttt{config\_store}. In contrast, the raw-alert diagnosis abandoned the hardware story only to construct a second false narrative, blaming an in-progress hypervisor image upgrade and disabled signature verification. Across five checkpoints from $t_0-80\text{h}$ through $t_0$, raw-alert prompting constructed two false narratives and never identified the actual defect, whereas the precedence forest reached a Full Diagnosis at $t_0-18\text{h}$ and kept it through $t_0$.

\paragraph{Case 2: Intermittent storage adapter disconnections and target path flapping.}
In this incident, the outage manifested at $t_0$ as severe storage access failures across the host. The root cause was command failures on a specific host bus adapter (HBA), which induced resource-lock exhaustion and memory-admission failures. Prior to $t_0$, the incubating fault generated recurring log alerts as a SAN target device repeatedly dropped and reconnected, incrementing its target identifier on every reconnect. At $t_0-26\text{h}$ (the earliest maximum-DQ checkpoint), \systemname{}'s precedence forest isolated this incubatory signature by linking the resource-lock and memory failures directly to recurring command errors on the affected adapter. This explicitly named the adapter and target device to achieve a Full Diagnosis 26 hours before $t_0$. In contrast, the LLM prompted with raw alerts over the same window independently detected path flapping. However, it attributed the alerts broadly to generic storage-network instability without pinpointing the specific adapter, yielding a Partial Diagnosis. At $t_0$, both representations maintained their relative accuracy: the precedence forest retained its Full Diagnosis on the specific adapter, while the raw-alert diagnosis continued to report broad storage instability.

\paragraph{Case 3: Recurring fabric driver memory allocation failure and command queue exhaustion.}
In this incident, the outage manifested at $t_0$ as total storage path loss and host isolation. The underlying cause was a third-party storage fabric driver that repeatedly failed to allocate an internal memory buffer and exhausted its command queue. Over six days prior to $t_0$, the incubating driver fault emitted sporadic warning alerts as storage paths dropped and degraded path redundancy. At $t_0-144\text{h}$ (the earliest maximum-DQ checkpoint), \systemname{}'s precedence forest linked a storage-queue-depth configuration alert to the adapter's throttling behavior. Its post-call subsystem-impact annotation isolated the storage-path configuration subsystem (37 descendant entities) to achieve a Partial Diagnosis 6 days before $t_0$. The raw-alert diagnosis at $t_0-144\text{h}$ and across six subsequent pre-incident checkpoints missed the fault entirely, misattributing alert volume to an unrelated storage-I/O accounting feature. Only at $t_0-2\text{h}$ did the raw-alert diagnosis detect the path resets. At $t_0$, the precedence forest connected the fabric driver's memory-allocation failure directly to the resulting storage disconnects, although this remained a Partial Diagnosis because the ground-truth entities were missing from the LLM response.

\paragraph{Case 4: Storage-volume quiesce-timeout notification registration fault.}
In this incident, the outage manifested at $t_0$ as cascading VM eviction failures and storage unmap errors during backup operations. The root cause was a storage volume failing to register for quiesce-timeout notifications, which freeze VM disk I/O during backup snapshots. A separate volume-format verification failure also occurred on an unrelated volume. Prior to $t_0$, background management operations generated sporadic warning alerts. At $t_0-86\text{h}$ (the earliest maximum-DQ checkpoint), \systemname{}'s precedence forest linked the volume-format verification alert to the ground-truth storage subsystem to achieve a Partial Diagnosis. However, its subsystem-impact annotation (30 subsystems across 613 entities) flagged the candidate as a broad shared dependency rather than an isolated root cause. In contrast, the LLM prompted with raw alerts identified only routine management activity and yielded no diagnostic signal. By $t_0-14\text{h}$ and at $t_0$, the signal was no longer retained as the forest changed across later checkpoints. Unlike the other case studies, in Case 4 an early partial signal did not persist through $t_0$.

\paragraph{Case 5: Session-scheduler configuration fault masked by coincidental network adapter flapping.}
In this incident, the outage manifested at $t_0$ as widespread management console unresponsiveness. The root cause was a missing host-side session-scheduler configuration file, which caused silent management-service refresh failures across ten independent user sessions. Prior to $t_0$, the missing file generated recurring background alerts across active sessions, while several physical network adapters coincidentally flapped during the same window. At $t_0-18\text{h}$ (the earliest maximum-DQ checkpoint), \systemname{}'s precedence forest linked the session-scheduler alerts recurring across all ten sessions. Its post-call subsystem-impact annotation isolated the host management-agent subsystem (55 entities) to achieve a Partial Diagnosis. In contrast, the LLM prompted with raw alerts at the identical pre-incident checkpoint constructed a false narrative. It attributed the pre-incident alerts to coincidental physical link flapping across the network adapters. At $t_0$, the precedence forest maintained its Partial Diagnosis based on session recurrence, whereas the raw-alert diagnosis persisted in its false narrative.

\subsection{Sensitivity and Ablations}
\label{sec:sensitivity}
To enable effective diagnosis, \systemname{} must fulfill three core operational objectives: (1)~maximize recall under aggressive search-space reduction to fit LLM context limits, (2)~suppress redundant alerts to mitigate operator fatigue, and (3)~construct interpretable branching structures that explicitly capture fault propagation. We evaluate parameter sensitivity to confirm robustness and systematically ablate key design choices across all $N=25$ enterprise incidents (Table~\ref{tab:ablation_final}) to quantify their individual contributions to these goals.

\noindent\textbf{Parameter Sensitivity.} We evaluated parameter sensitivity across correlation window $T_c$, onset window $T_w$, and session walk cap $K_{\text{walk}}$ using a 5-incident enterprise subset, reported next. The sweeps confirm that our defaults ($T_c=2.0\text{h}$, $T_w=2.0\text{h}$, $K_{\text{walk}}=15$) achieve reasonable trade-offs: $T_c$ bounds edge expansion, $T_w$ balances onset responsiveness against alert volume, and $K_{\text{walk}}$ prevents alert fragmentation by allowing transitive closures to reach natural completion. Open-source benchmarks remain unaffected because their shorter execution timelines fit comfortably within these bounds.

Tables~\ref{tab:tc_sensitivity}--\ref{tab:kwalk_sensitivity} report these sweeps. All metrics report mean values across the 5 bundles. In addition to standard recall and reachability (\S\ref{sec:recall}), we track average forest size (\textit{Mean Nodes}) and alert volume (\textit{Alerts/Incident}). While $T_w$ and $K_{\text{walk}}$ directly govern progressive alert generation, $T_c$ strictly bounds edge formation within the bipartite graph.

As correlation window $T_c$ increases, events that are distant in time become eligible for inclusion in the bipartite entity-event graph. As a result, larger forests are formed as shown in Table~\ref{tab:tc_sensitivity}; overall recall remains constant ($0.87$) while Recall@1600 increases slightly ($0.61 \to 0.66$).

\begin{table}[t]
\centering
\scriptsize
\caption{\textbf{Sensitivity to Correlation Window ($T_c$).} Evaluated across 5 enterprise incidents keeping $T_w=2.0\text{h}$ and $K_{\mathrm{walk}}=15$ constant. Increasing $T_c$ expands average forest node size while yielding modest gains in prompt-capped recall (Recall@1600).}
\label{tab:tc_sensitivity}
{\setlength{\tabcolsep}{2pt}
\begin{tabular}{@{}p{2.5cm}>{\centering\arraybackslash}p{0.85cm}>{\centering\arraybackslash}p{1.35cm}>{\centering\arraybackslash}p{1.55cm}>{\centering\arraybackslash}p{1.05cm}@{}}
\toprule
\textbf{Correlation Window ($T_c$)} & \textbf{Top-10} & \textbf{Recall@All} & \textbf{Recall@1600} & \textbf{Mean Nodes} \\ \midrule
$1.0\text{h}$                       & 60.0\%          & 0.87                 & 0.61                  & 11,426.5            \\
$2.0\text{h}$ (Default)             & \textbf{80.0\%} & 0.87                 & 0.61                  & 11,979.7            \\
$8.0\text{h}$                       & \textbf{80.0\%} & 0.87                 & \textbf{0.66}         & 14,000.7            \\
$24.0\text{h}$                      & \textbf{80.0\%} & 0.87                 & \textbf{0.66}         & 14,328.1            \\ \bottomrule
\end{tabular}
}
\end{table}

On the other hand, the onset window $T_w$ trades off fine-grained onset detection against alert volume (Table~\ref{tab:tw_sensitivity}). Smaller $T_w$ values allow failure onset to be detected early, whereas larger $T_w$ values combine more unrelated events into the same forest, reducing Top-10 reachability ($80.0\% \to 60.0\%$). While $T_w=1.0\text{h}$ achieves identical recall to $T_w=2.0\text{h}$, it nearly doubles alert generation ($48.4$ vs.\ $26.2$ alerts/incident), making $T_w=2.0\text{h}$ the practical default.

\begin{table}[t]
\centering
\scriptsize
\caption{\textbf{Sensitivity to Window Size ($T_w$).} Evaluated across 5 enterprise incidents keeping $T_c=2.0\text{h}$ and $K_{\mathrm{walk}}=15$ constant. Expanding $T_w$ dramatically suppresses alert volume, but larger windows over-aggregate events, expanding forest size and degrading reachability.}
\label{tab:tw_sensitivity}
{\setlength{\tabcolsep}{2pt}
\begin{tabular}{@{}p{1.4cm}>{\centering\arraybackslash}p{0.8cm}>{\centering\arraybackslash}p{1.35cm}>{\centering\arraybackslash}p{1.55cm}>{\centering\arraybackslash}p{1.2cm}>{\centering\arraybackslash}p{1.2cm}@{}}
\toprule
\textbf{Window Size ($T_w$)} & \textbf{Top-10} & \textbf{Recall@All} & \textbf{Recall@1600} & \textbf{Mean Nodes} & \textbf{Alerts / Incident} \\ \midrule
$1.0\text{h}$                & \textbf{80.0\%} & \textbf{0.87}        & \textbf{0.61}         & \textbf{11,182.9}   & 48.4                          \\
$2.0\text{h}$ (Default)      & \textbf{80.0\%} & \textbf{0.87}        & \textbf{0.61}         & 11,979.7            & 26.2                          \\
$8.0\text{h}$                & 60.0\%          & 0.85                 & \textbf{0.61}         & 17,526.7            & 8.4                           \\
$16.0\text{h}$               & 60.0\%          & 0.85                 & 0.56                  & 17,791.3            & \textbf{5.4}                  \\ \bottomrule
\end{tabular}
}
\end{table}

Finally, the walk cap $K_{\mathrm{walk}}$ controls maximum depth during progressive onset tracking. As a failure spreads across time steps, the walk continues until the transitive closure naturally terminates—meaning no new correlated events appear. If $K_{\mathrm{walk}}$ is too low, the walk is forcibly truncated and restarted in a new session later, creating fragmented alerts. Raising $K_{\mathrm{walk}}$ allows walks to reach natural completion, expanding forest size while consolidating related activity into fewer alerts ($31.6 \to 26.0$ alerts/incident) and improving overall recall ($0.87 \to 0.90$ at $K_{\mathrm{walk}}=25$, Table~\ref{tab:kwalk_sensitivity}). While these defaults ($T_c=2.0\text{h}, T_w=2.0\text{h}, K_{\mathrm{walk}}=15$) serve as baseline thresholds for batch evaluation, production deployments operating over continuous streaming logs must tune these bounds based on real-time stream latency and alert-volume budgets.

\begin{table}[t]
\centering
\scriptsize
\caption{\textbf{Sensitivity to Walk Cap ($K_{\mathrm{walk}}$).} Evaluated across 5 enterprise incidents keeping $T_c=2.0\text{h}$ and $T_w=2.0\text{h}$ constant. Extending walk cap allows transitive closure to naturally converge, consolidating events into fewer alerts and improving overall recall.}
\label{tab:kwalk_sensitivity}
{\setlength{\tabcolsep}{2pt}
\begin{tabular}{@{}p{1.4cm}>{\centering\arraybackslash}p{0.8cm}>{\centering\arraybackslash}p{1.35cm}>{\centering\arraybackslash}p{1.55cm}>{\centering\arraybackslash}p{1.2cm}>{\centering\arraybackslash}p{1.2cm}@{}}
\toprule
\textbf{$K_{\mathrm{walk}}$ Cap} & \textbf{Top-10} & \textbf{Recall@All} & \textbf{Recall@1600} & \textbf{Mean Nodes} & \textbf{Alerts / Incident} \\ \midrule
5 steps                     & 80.0\%          & 0.87                 & 0.61                  & \textbf{9,334.3}    & 31.6                          \\
\textbf{15 steps (Default)} & 80.0\%          & 0.87                 & 0.61                  & 11,979.7            & 26.2                          \\
25 steps                    & 80.0\%          & \textbf{0.90}        & 0.61                  & 12,765.3            & \textbf{26.0}                 \\ \bottomrule
\end{tabular}
}
\end{table}

\noindent\textbf{Architectural Ablations.} Disabling individual design choices across all $N=25$ enterprise incidents demonstrates their necessity (Table~\ref{tab:ablation_final}):

\begin{table*}[t]
\centering
\scriptsize
\caption{\textbf{Ablation Study Summary.} Impact of each design decision on search space reduction, recall, and diagnostic performance.}
\label{tab:ablation_final}
\begin{tabular}{@{}p{3.3cm}p{2.2cm}p{2.4cm}>{\centering\arraybackslash}p{1.2cm}>{\centering\arraybackslash}p{1.2cm}p{2.2cm}p{2.4cm}@{}}
\toprule
\textbf{Design Decision} & \textbf{Ablation Variant} & \textbf{Search Space Reduction ($N{=}25$)} & \textbf{Recall@All} & \textbf{Recall@1600} & \textbf{Top-10 / DQ} & \textbf{Primary Impact} \\ \midrule
Bounded transitive closure ($T_c$) & Production ($T_c{=}2\text{h}$) & Events: 85.7\% med / 69.5\% mean & 0.76 & 0.46 & 60.0\% & Search space reduction only \\ \addlinespace
Severity elevation & Real $\to$ No subsystem promotion & N/A & 0.96 $\to$ 0.49 & N/A & N/A & Recall critical \\ \addlinespace
Progressive onset tracking & Flat 7-day $\to$ Windowed & N/A & 0.59 $\to$ 0.76 & 0.31 $\to$ 0.46 & 52.0\% $\to$ 60.0\% & Recall \& reachability gain \\ \addlinespace
Hub-node pruning & ON $\to$ OFF & Entities: 2.8\% med / 14.9\% mean & 0.76 $\to$ 0.77 & 0.46 $\to$ 0.46 & DQ (Full): 10/25 $\to$ 8/25 & Search space reduction; modest DQ effect \\ \addlinespace
Parent-edge assignment rule & Nearest $\to$ Earliest & Graph structure only & 0.76 $\to$ 0.76 & 0.46 $\to$ 0.46 & 24.2\% $\to$ 97.5\% leaf nodes & Recall neutral; graph interpretability \\ \addlinespace
Forest structural tagging & OFF $\to$ ON & Graph structure only & 0.76 $\to$ 0.76 & 0.46 $\to$ 0.46 & DQ (Full): 5/25 $\to$ 10/25 & Recall neutral; large DQ gain \\ \bottomrule
\end{tabular}
\end{table*}

\noindent Bounded transitive closure drives search-space reduction: a $T_c = 2\text{~h}$ window filters a median of $85.7\%$ (mean $69.5\%$) of background events, preserving $0.76$ mean recall. We do not evaluate a fully unbounded ($T_c=\infty$) variant: unlike the memory exhaustion that caps graph baselines (\S\ref{sec:background}), an unbounded window instead risks merging temporally disjoint but structurally connected events into one component, polluting the cascade with unrelated history.
Widening $T_c$ from $1\text{~h}$ to $24\text{~h}$ (Table~\ref{tab:tc_sensitivity} above) yields only modest Recall@1600 gains ($0.61\to0.66$) while continuing to expand forest size.

\noindent Severity elevation is critical for recall; disabling its subsystem-level severity promotion (one of its two mechanisms) causes recall to drop from 0.96 to 0.49, and disabling its log-text keyword override alone drops it to 0.65 (both measured on the raw 168h candidate pool across all $N=25$ incidents, counting the one incident whose ground truth is not isolated as zero recall), because true faults that appear to be benign are dropped before graph construction.

\noindent Progressive onset tracking ($T_w = 2\text{h}$) replaces a flat 7-day log sweep and boosts overall recall from 0.59 to 0.76 (0.31 to 0.46 capped) and increases Top-10 reachability from 52.0\% to 60.0\% by isolating the true onset window.

\noindent Hub-node pruning removes ubiquitous hub entities and reduces the entity search space by a median of 2.8\% (mean 14.9\%); disabling pruning slightly increases raw graph recall ($0.76$ to $0.77$), but it modestly degrades LLM DQ (Full Diagnoses drop from 10/25 to 8/25) because unpruned hub nodes appear as spurious root candidates. The current prompt's explicit per-candidate semantic fault check makes the model robust to unpruned hub noise, though pruning still helps at the margin.

\noindent The parent-edge assignment rule is recall-invariant in our measurements ($0.76$ uncapped / $0.46$ capped), but nearest-preceding parent selection yields a far more informative graph structure than earliest-preceding selection. Under earliest-preceding logic, the initial root entity acts as a hub, forcing intermediate faults to attach as direct leaves; the fraction of leaf nodes rises from 24.2\% to 97.5\%, losing propagation information.

\noindent Forest structural tagging, which adds structural tags (\verb+[ROOT_CANDIDATE]+ and \verb+[DEAD_END]+) to the prompt, is recall-neutral ($0.76$ / $0.46$) and produces a large DQ gain under the current prompt (Full Diagnoses rise from 5/25 to 10/25, while Full + Partial rises from 15/25 to 20/25; Partial Diagnoses alone hold flat at 10/25, since the gain is a mix of Misses converting directly to Full Diagnoses and borderline Partial Diagnoses sharpening into Full ones), confirming that tags meaningfully guide LLM reasoning for this template.

\subsection{Compute and Memory Overheads}
\label{sec:overheads}

\begin{figure*}[t]
\centering
\includegraphics[width=\linewidth]{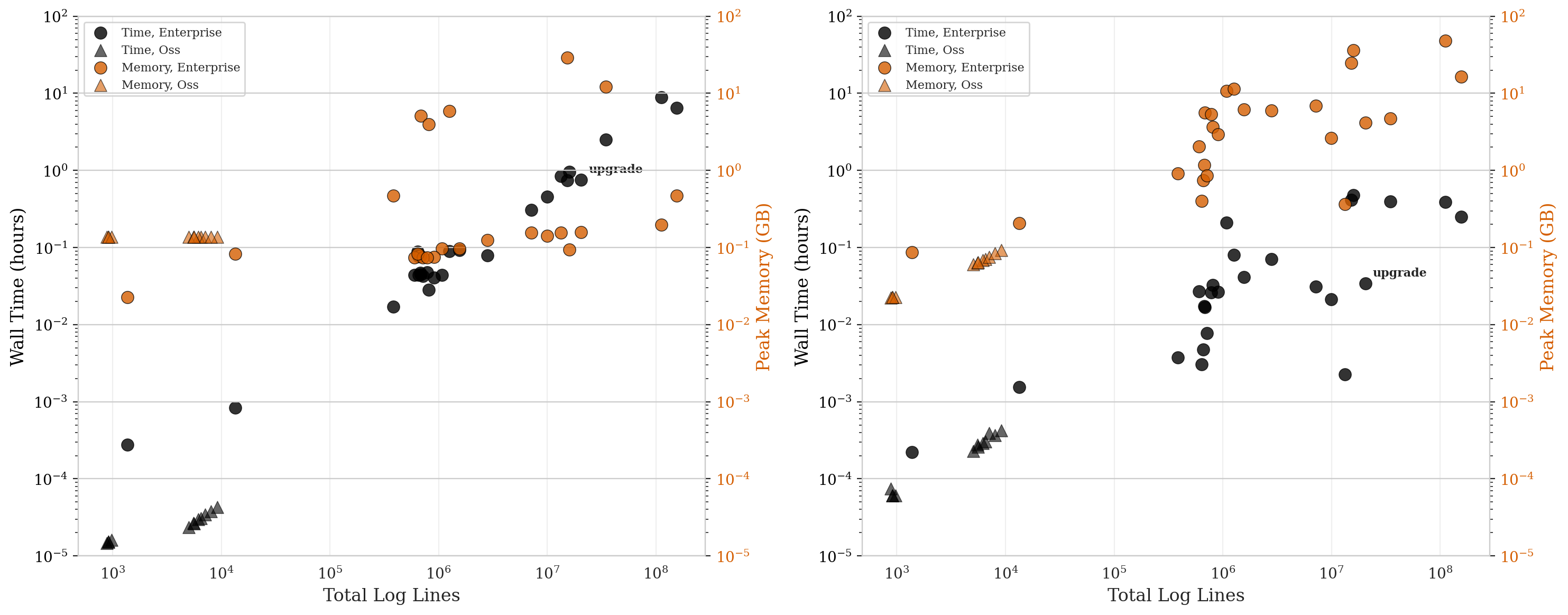}
\caption{\textbf{Compute and memory overheads, split by pipeline stage.} Both panels plot wall-time in hours (black, left $y$-axis) and peak memory in GB (orange, right $y$-axis) against total raw log lines $N$ on a log-log scale across all 37 benchmark datasets (25 enterprise circles, 12 open-source triangles). \emph{Left}: Log Ingestion, where wall-time scales linearly with $N$ ($r=0.99$) while peak memory remains bounded ($18.7\text{--}479.9$\,MB) due to fixed-batch streaming. \emph{Right}: Graph Construction, where overheads correlate with input scale $N$ through window exploration volume ($r=0.92\text{--}0.93$), reaching a peak memory ceiling of 48.05\,GB in the single worst case.}
\label{fig:hardware_metrics_split}
\end{figure*}

To evaluate computational overheads, we measured processing wall-time (in hours) and peak memory usage (in gigabytes) across all 37 benchmark datasets. Figure~\ref{fig:hardware_metrics_split} illustrates performance across the two pipeline stages plotted against total raw log lines $N$.

The left panel shows log ingestion overheads. Execution wall-time scales linearly with $N$ ($r = 0.99$) because ingestion performs a single-pass parsing step over each raw log line. Enterprise datasets exhibit a median ingestion time of 2.9\,minutes (0.4--5.4\,seconds for open-source datasets), reaching a maximum of 6.4\,hours on the largest dataset ($156.4\text{M}$ lines). Peak memory stays strictly bounded between 18.7\,MB and 479.9\,MB because logs are processed in fixed-size batches rather than loaded into memory all at once.

The right panel shows graph construction overheads. Both wall-time and peak memory correlate closely with input scale $N$ ($r = 0.92\text{--}0.93$) because progressive onset tracking evaluates multiple candidate onset windows across the log corpus. For enterprise datasets, median graph construction wall-time is 1.6\,minutes (0.96\,seconds for open-source) and peaks at 28.9\,minutes on the largest corpus. Peak memory reaches 4.17\,GB at the median and caps at 48.05\,GB in the single worst case, remaining well within host memory capacity.

%% file: sections/related.tex
\section{Related Work}
\label{sec:related}
Over the past two decades, a rich body of diagnostic research has addressed failure localization. In the context of this work, prior studies broadly fall into two main paradigms: 

(1) \textbf{Domain-Augmented Methods}, which rely on pre-existing topology maps or CMDBs~\cite{nadi2009draca, skynet, arzani2018007, kakarla2024diffy}, historical incident tickets and runbooks~\cite{chen2024rcacopilot, dogga2022revelio, jiang2020mitigate, shetty2022autotsg}, active instrumentation~\cite{sigelman2010dapper, alvaro_ldfi, zhang_csnake26, yuan2010sherlog}, or graph-based ranking over metric/trace topologies~\cite{kandula2009netmedic, kim2013root, chen2014causeinfer, wu2020microrca}; and

(2) \textbf{Statistical \& Performance Inference Methods}, which infer causal dependencies or isolate execution anomalies using log sequence probabilities, statistical delay distributions, or transaction success rates~\cite{xu2009detecting, du2017deeplog, guo2021logbert, meng2019loganomaly, bahl2007sherlock, aguilera2003performance}.

\systemname{} occupies the opposite extreme: it bounds the search space in black-box environments where neither domain artifacts, historical incident corpora, active instrumentation, nor healthy operational baselines exist. In this sense, its underlying objective is distinct: it investigates the extent to which noisy log data alone can encode failure propagation.

\noindent\textbf{Domain-Augmented Methods.}
Among diagnostic paradigms that exploit domain artifacts, \systemname{} resembles graph-based inference algorithms~\cite{kandula2009netmedic, kim2013root, chen2014causeinfer, wu2020microrca} in that it relies on the premise that failures propagate along structural dependencies best captured as a directed graph. It differs from them by deriving these structural dependencies entirely post-hoc from noisy log data rather than relying on high-fidelity time-series metrics, active network probing, or pre-existing topology maps.

\systemname{} is also conceptually inspired by distributed tracing and request-reconstruction frameworks~\cite{barham2004magpie, fonseca2007xtrace, sigelman2010dapper, mace2015pivot, kaldor2017canopy}. While explicit tracing systems (e.g., Dapper~\cite{sigelman2010dapper}, X-Trace~\cite{fonseca2007xtrace}, Canopy~\cite{kaldor2017canopy}) require unbroken \texttt{TraceID} header propagation, and schema-driven tools (e.g., Magpie~\cite{barham2004magpie}) depend on pre-authored event schemas to join local handles, \systemname{} operates on unmodeled, local identifiers (e.g., task handles, disk IDs) that are frequently noisy, fragmented, or missing. \systemname{} overcomes this fragmentation without schemas or headers by exploiting identifier co-occurrence across intermediate log events to stitch failure paths together transitively across subsystem boundaries. 

Other diagnostic frameworks rely on explicit request instrumentation, static code analysis, fault injection, or pre-built network topology maps~\cite{sigelman2010dapper, yuan2010sherlog, zhang_csnake26, nadi2009draca, bittner2016tfpg, alvaro_ldfi, arzani2018007}. Most closely related in intent to our alert-reduction mechanism is SkyNet~\cite{skynet}, which aggregates alert floods from passive monitoring telemetry; however, SkyNet requires known network topology and hierarchy maps to group alerts, whereas \systemname{} performs event correlation completely independent of system topology.

Finally, \systemname{} intersects with recent generative AI and agentic AIOps frameworks~\cite{jiang2020mitigate, shetty2022autotsg, chen2024rcacopilot, wang2024rcaagent, zhang2025tamo, xu2026aida, ren2026xihe}. Like AIDA~\cite{xu2026aida} and Xihe~\cite{ren2026xihe}, \systemname{} recognizes that failure propagation is best encoded as a directed graph rather than flat entity lists. However, unlike frameworks requiring pre-authored playbooks~\cite{jiang2020mitigate, shetty2022autotsg} or historical ticket corpora~\cite{chen2024rcacopilot, dogga2022revelio}, \systemname{} operates with zero historical domain knowledge. \systemname{} complements these frameworks by converting raw log streams into compact precedence forests to provide bounded context for LLMs when playbooks do not exist and raw logs exceed context windows.

We do not benchmark directly against agentic LLM-RCA systems: each relies on a closed, organization-specific incident corpus with no public implementation, and faithfully reproducing them would require curated incident-handler libraries and historical ticket archives—artifacts whose absence is this paper's core premise. Synthesizing such resources would test a different, better-instrumented regime than the zero-prior-knowledge environment \systemname{} targets.

\noindent\textbf{Statistical \& Performance Inference Methods.}
Within the second paradigm, \systemname{} is similar to log-only sequence models~\cite{xu2009detecting, du2017deeplog, guo2021logbert, meng2019loganomaly, li2020swisslog, zhang2019logrobust, zhang2021onion, jia2023rtlog} in that it abstracts raw log text into parameterized event templates. It is also conceptually aligned with black-box performance inference frameworks like Aguilera et al.~\cite{aguilera2003performance} and Sherlock~\cite{bahl2007sherlock}, which infer component dependencies from passive telemetry without requiring source code or explicit schemas.

\systemname{} differs in two ways: it assigns precedence edges directly from single instances of entity co-occurrence (individual log lines carry more than one entity, $N_{\text{line}} > 1$; Table~\ref{tab:entity_density}) rather than statistical aggregation across thousands of repeated transactions, and it explicitly leverages dynamic entity identifiers as relational glue rather than masking them as wildcards~\cite{li2023didmissimportantstudying, he2022empirical}.

%% file: sections/limits.tex
\section{Limitations and Open Problems}
\label{sec:limitations}
\noindent\textbf{Interactivity.} \systemname{} executes as a single batch pass over a log bundle, without iterative refinement or operator-guided narrowing mid-investigation. It provides the initial bounded context an interactive agent or human operator begins from, rather than replacing interactive triage itself.

\noindent\textbf{Log Parsing.} \systemname{} abstracts log text into parameterized templates using lightweight vendor integration modules (\S\ref{sec:implementation}), following standard log abstraction paradigms~\cite{he2017drain, liu2022uniparser, le2023logparsing, zhang2023system}. This module adds a one-time, per-vendor engineering cost. Incomplete parameter masking can misclassify identical logs as distinct events, inducing template fragmentation that transitive closure overcomes via shared entities. Separately, an uncaptured log-level keyword in 1 of 25 incidents prevented severity elevation from isolating its ground-truth fault, so ground truth is detectable in only $24/25$ incidents; this caps mean enterprise recall at $0.96$.

\noindent\textbf{Maximal Cascade Selection.} Isolating the maximal connected component ($F^*$) maximizes search-space reduction but causes zero recall in 3 of 25 incidents whose root causes resided in secondary subgraphs (distinct from the 1 of 25 severity-elevation miss above; the $0.76$ mean recall defined in \S\ref{sec:recall} additionally reflects partial recall within incidents); emitting all subgraphs recovers these cases while still yielding a $2.2\times$ net alert reduction.

\noindent\textbf{Ground-Truth and Scoring Boundaries.} Ground-truth entities and subsystems were reconstructed directly from vendor incident reports and escalation logs rather than synthetic labeling, reflecting real-world operational diagnoses. Our scoring harness enforces strict entity identifier matching for Full Diagnoses to prevent substring false positives; consequently, a response that correctly identifies the root failure mechanism without citing specific entity tokens is conservatively scored as a Partial Diagnosis.

\noindent\textbf{Third-Party Subsystem Boundary.} Third-party vendor software represents the engine's primary visibility limit: while \systemname{} achieves $100\%$ accuracy ($9/9$) on native infrastructure faults, accuracy drops to $69\%$ ($11/16$) on third-party components, accounting for all $5$ diagnostic misses. Custom per-vendor parsing modules and targeted severity elevation serve as the primary engineering levers to lift this floor in heterogeneous enterprise deployments.

\noindent\textbf{The Gap to Self-Healing.} \systemname{} exploits structural and temporal correlation, not causality; leveraging it for automated remediation requires domain reasoning to bridge this gap before executing corrective actions.

\noindent\textbf{Telemetry Gaps.} When network hardware lacks text logging or adjacent domains use disjoint identifier schemes, transitive closure cannot complete and the precedence forest fractures at the unlinked boundary; silent infrastructure failures remain unrecoverable from logs alone, requiring multi-modal telemetry (topology maps, metric anomalies) to bridge these gaps.

%% file: sections/conc.tex
\section{Conclusion}
\label{sec:conc}

This work shows that log-only failure diagnosis holds strong promise in enterprise domains with rich logging. By eliminating a median $99.8\%$ of background noise in a median wall-time of 4.5~minutes while retaining $0.76$ mean recall, \systemname{} demonstrates that aggressive search-space reduction is achievable on raw text logs alone. Operating on 2-hour log windows, its 16-hour median diagnosis-verified lead time and $124\times$ consolidation into structured, fault-linking alerts establish a viable design point for AIOps: rather than requiring pre-built domain artifacts or historical baselines, its compressed failure graphs support both zero-shot LLM root-cause identification ($80\%$ enterprise, $100\%$ open-source; \S\ref{sec:interpret}) and direct operator triage. \systemname{} is currently being integrated into the AIOps pipeline of an enterprise cloud provider to automate incident triage and accelerate ticket resolution.